\pdfoutput=1
\documentclass[twocolumn,prd,superscriptaddress,preprintnumbers,nofootinbib]{revtex4}
\usepackage{graphicx}
\usepackage{epsfig}
\usepackage[rightcaption]{sidecap}
\usepackage{lipsum}
\usepackage{enumitem}
\usepackage{bm}
\usepackage{latexsym,amssymb,amsmath,amsfonts,amssymb,txfonts,pxfonts,wasysym,float}
\usepackage{color}

\newcommand{\beq}[1]{\begin{equation}\label{#1}}
\newcommand{\eeq}{\end{equation}}
\newcommand{\bea}[1]{\begin{eqnarray} \label{#1}}
\newcommand{\eea}{\end{eqnarray}}
\newcommand{\ba}{\begin{array}}
\newcommand{\ea}{\end{array}}

\def\be{\begin{equation}}
\def\ee{\end{equation}}
\def\gs{\mathrel{
   \rlap{\raise 0.511ex \hbox{$>$}}{\lower 0.511ex \hbox{$\sim$}}}}
\def\ls{\mathrel{
   \rlap{\raise 0.511ex \hbox{$<$}}{\lower 0.511ex \hbox{$\sim$}}}}

\newcommand{\comment}[1]{}

\usepackage[usenames,dvipsnames]{xcolor}
\definecolor{orange}{cmyk}{0,0.5,1,0}
\definecolor{rossoCP3}{cmyk}{0,.88,.77,.40}
\definecolor{graa}{rgb}{0.8,0.8,0.8}
\definecolor{blaa}{rgb}{0.2,0.2,0.6}

\RequirePackage[normalem]{ulem}

\begin{document}
\preprint{DESY-22-021}

\title{\color{rossoCP3}{%
An explanation of the muon puzzle of ultrahigh-energy cosmic rays\\
and the role of the Forward Physics Facility for model improvement}}

\author{Luis A. Anchordoqui}

\affiliation{Department of Physics and Astronomy,  Lehman College, City University of
  New York, NY 10468, USA
}

\affiliation{Department of Physics,
 Graduate Center, City University
  of New York,  NY 10016, USA
}

\affiliation{Department of Astrophysics,
 American Museum of Natural History, NY
 10024, USA
}

\author{Carlos Garc\'{\i}a Canal}

\affiliation{Instituto de
  F\'{\i}sica La Plata - CONICET and Departamento de F\'{\i}sica,
  Facultad de Ciencias Exactas,\\ Universidad Nacional de La Plata,
  C.C. 69, (1900) La Plata, Argentina}

\author{Felix Kling}
\affiliation{Deutsches Elektronen-Synchrotron DESY, Notkestr. 85, 22607 Hamburg, Germany}

\author{Sergio J. Sciutto}

\affiliation{Instituto de
  F\'{\i}sica La Plata - CONICET and Departamento de F\'{\i}sica,
  Facultad de Ciencias Exactas,\\ Universidad Nacional de La Plata,
  C.C. 69, (1900) La Plata, Argentina}

\author{Jorge F. Soriano}

\affiliation{Department of Physics and Astronomy,  Lehman College, City University of
  New York, NY 10468, USA
}

\begin{abstract}

  \vskip 2mm \noindent We investigate the observed muon deficit in
  air shower simulations when compared to ultrahigh-energy cosmic ray
  (UHECR) data.  Based upon the observed enhancement of strangeness
  production in high-energy hadronic collisions reported by the ALICE
  Collaboration, the concomitant $\pi \leftrightarrow K$ swap is
  considered as the keystone to resolve the muon anomaly through its
  corresponding impact on the shower development.  We construct a toy
  model in terms of the $\pi \leftrightarrow K$ swapping probability
  $F_s$. We present a parametrization of $F_s$ in terms of the
  pseudorapidity that can accommodate the UHECR data.  Looking to the
  future, we explore potential strategies for model improvement using
  the massive amounts of data to be collected by LHC neutrino
  detectors, such as FASER$\nu$ and experiments at the Forward Physics
  Facility. We calculate the corresponding sensitivity to $F_s$ and show
  that these experiments
  will be able to probe the model phase space.
\end{abstract}
\maketitle

\section{Introduction}

Ultra-high-energy ($10^9 \lesssim E/{\rm GeV} \lesssim 10^{11}$)
cosmic ray (UHECR) collisions have center-of-mass energies
($50 \lesssim \sqrt{s}/{\rm TeV} \lesssim 450$) well beyond those
achieved at collider experiments, and thereby provide an invaluable
probe of particle interactions below the fermi
distance~\cite{Anchordoqui:2018qom}. Of particular interest here, the
highest energy cosmic rays currently observed by the Pierre Auger
Observatory (Auger)~\cite{PierreAuger:2015eyc,Aab:2014pza,Aab:2016hkv} and the Telescope
Array~\cite{TelescopeArray:2012uws,Tokuno:2012mi,TelescopeArray:2018eph} show a significant discrepancy in
the shower muon content when compared to predictions of LHC-tuned
hadronic event generators~\cite{dEnterria:2011twh}. More concretely,
the analysis of Auger data suggests that the hadronic component of
showers (with primary energy $10^{9.8} < E/{\rm GeV} < 10^{10.2}$)
contains about $30\%$ to $60\%$ more muons than expected. The
significance of the discrepancy between Auger data and model
prediction is somewhat above $2.1\sigma$~\cite{Aab:2016hkv}.  Auger
findings have been recently confirmed studying air shower
measurements over a wide range of energies. The muon deficit between
simulation and data, dubbed the {\it muon puzzle}, starts at $E \sim
10^8~{\rm GeV}$ increasing noticeably as primary energy grows, with a
slope which was found to be
significant at about $8\sigma$~\cite{EAS-MSU:2019kmv,Cazon:2020zhx,Dembinski:2021szp}.

Certainly, in solving the muon puzzle one has to simultaneously get a
good agreement with the measurements of the distribution of the depth
of shower maximum $X_{\rm max}$~\cite{PierreAuger:2014sui}, and the fluctuations in the number
of muons~\cite{PierreAuger:2021qsd}. A thorough phenomenological study
has shown that an unrivaled solution to the muon deficit, compatible
with the observed $X_{\rm max}$ distributions, is to reduce the
transfer of energy from the hadronic shower into the electromagnetic
shower, by reducing the production or decay of neutral
pions~\cite{Allen:2013hfa}. Several models have been proposed to
accommodate this effect, including those wherein strangeness
production suppresses the pion-to-kaon
ratio~\cite{Farrar:2013lra,Anchordoqui:2016oxy,Baur:2019cpv}. This
modification could have a compounded effect on the hadronic cascade,
so that only a comparably small reduction of $\pi^0$ production is
required. 

We note in passing that the proposed enhancement of
strangeness production in high-energy hadronic collisions was observed
by ALICE in the mid-rapidity
region~\cite{ALICE:2016fzo}. Specifically, ALICE observations show an
enhancement of the yield ratio of strange and multi-strange hadrons to
charged pions as a function of multiplicity at mid-rapidity not only
in PbPb and XeXe collisions but also in $pp$ and $p$Pb
scattering~\cite{Palni:2019ckt}. It goes without saying that none of
the hadronic interaction models currently used in air shower
simulations correctly reproduce ALICE
data~\cite{Anchordoqui:2019laz}. Assuming that the observed
enhancement of strangeness production in high-energy hadronic
collisions is at the core of the muon puzzle in this paper we study
the concomitant $\pi \leftrightarrow K$ swap impact on the development
of extensive air showers (EASs), using  phenomenological toy models
implemented in AIRES (version 19.04.08)~\cite{Sciutto:1999jh}. After
that, we discuss the prospects to constrain our model using
forward neutrino flux measurements at
FASER$\nu$~\cite{FASER:2019dxq,FASER:2020gpr} and future experiments
at the Forward Physics Facility (FPF)~\cite{Anchordoqui:2021ghd}.

There are two points
worth noting at this juncture: {\it (i)} The mid-rapidity region in
which the ALICE Collaboration reported a universal strangeness
enhancement in $pp$, $p$Pb and PbPb collisions is not directly
relevant for air showers experiments. It has not been observed
experimentally yet whether these effects could also be seen in hadrons
produced at forward rapidities. This is the {\it main assumption} of
our model, which will be directly tested at the FPF. {\it
(ii)}~Accommodating the muon deficit between simulations and data can
be virtually reduced to a constant factor, which is independent of the
primary energy~\cite{Sciutto:2019pqs}. In our toy model this factor is
taken to be related to the $\pi \leftrightarrow K$ swapping
probability.

The layout of the paper is as follows. In Sec.~\ref{sec:2} we first
discuss general aspects of a toy model and describe the (input and
output) AIRES module interface. Armed with the new AIRES module we
confront the toy model with Auger data. We perform a parameter scan
using results of EAS simulations and determine the phase space
boundaries of the $\pi \leftrightarrow K$ swapping probability from
experimental data. In Sec.~\ref{sec:3} we improve our toy model to
transform it into a predictive model.  We present a parametrization of
the $\pi \leftrightarrow K$ swapping probability in terms of the
pseudorapidity that can accommodate the UHECR data. In Sec.~\ref{sec4}
we investigate the sensitivity to the $\pi \leftrightarrow K$
swapping probability at FASER$\nu$ and the FPF and demonstrate that a direct test of the model
predictions is indeed feasible. The paper wraps up with some
conclusions presented in Sec.~\ref{sec5}.

\section{A Toy Model}
\label{sec:2}
To describe the shower evolution we adopt the AIRES simulation engine~\cite{Sciutto:1999jh} which provides full
space-time particle propagation in a realistic environment. The features of the AIRES version used for this work (19.04.08) are explained in detail in Ref.~\cite{Sciutto:1999jh}. 

For the present analysis, we prepared a new module to account for the
possible enhancement of strangeness production in high-energy hadronic
collisions. Every time an hadronic collision is processed, the list of
secondary particles obtained from the external  event generator 
invoked (for our analysis we adopt SIBYLL 2.3d~\cite{Riehn:2019jet}) is scanned by the new module before passing it to the main particle propagating engine.  The main
characteristics of the new AIRES module are as follows.

\subsection{Model Parameters}
\begin{widetext}

\begin{center}
\begin{tabular}{p{0.25\textwidth}cp{0.60\textwidth}}
{\bf Swapping fraction} \dotfill     & $f_s$ &
     Controls the kind and number of secondary particles that are
     affected by change of identity: $-1\le f_s \le 1$. In this zeroth-order approximation we take the swapping probability $F_s = f_s$.
\\*[10pt]
{\bf Projectile energy range} \dotfill & $[E_{\rm pmin},E_{\rm pmax}]$ &
     Particle swapping is performed only in hadronic collisions where the
     projectile kinetic energy verifies
     $E_{\rm pmin}\le E_{\rm proj}<E_{\rm pmax}$. $E_{\rm pmin}$ must
     be larger than 900 MeV and less than $E_{\rm pmax}$.
     We set $E_{\rm pmax}\to\infty$ unless otherwise specified.
\\*[10pt]
{\bf Secondary energy range} \dotfill & $[E_{\rm smin},E_{\rm smax}]$ &
     Secondary particles with kinetic energies out of the range
     $[E_{\rm smin},E_{\rm smax}]$ are always left unchanged.
     $E_{\rm smin}$ must be larger than 600 MeV and less than $E_{\rm smax}$.
     We set $E_{\rm smin} = 1~{\rm TeV}$, and $E_{\rm smax}\to\infty$
     unless otherwise specified.
\end{tabular}
\end{center}
\end{widetext}

\subsection{Logics of Hadronic Collision Post-Processing}
\label{sec:2b}

\begin{figure*}
\centering
\begin{tabular}{ccccc}
    \includegraphics[width=0.30\textwidth]{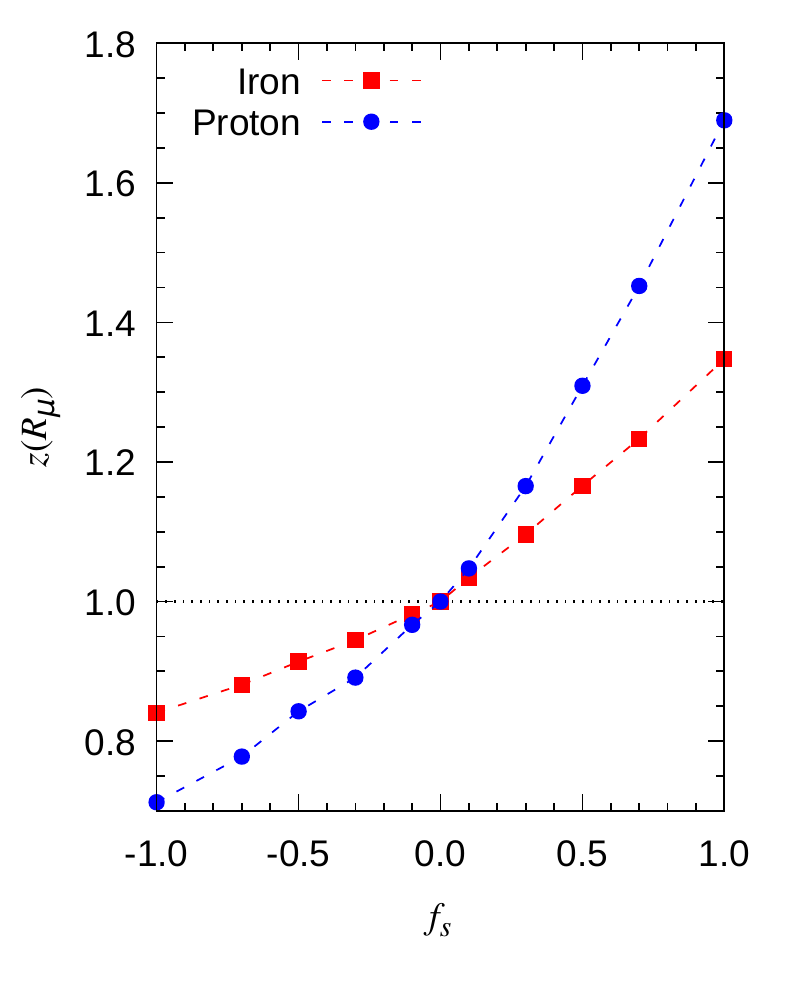}&
    \kern1em &
    \includegraphics[width=0.30\textwidth]{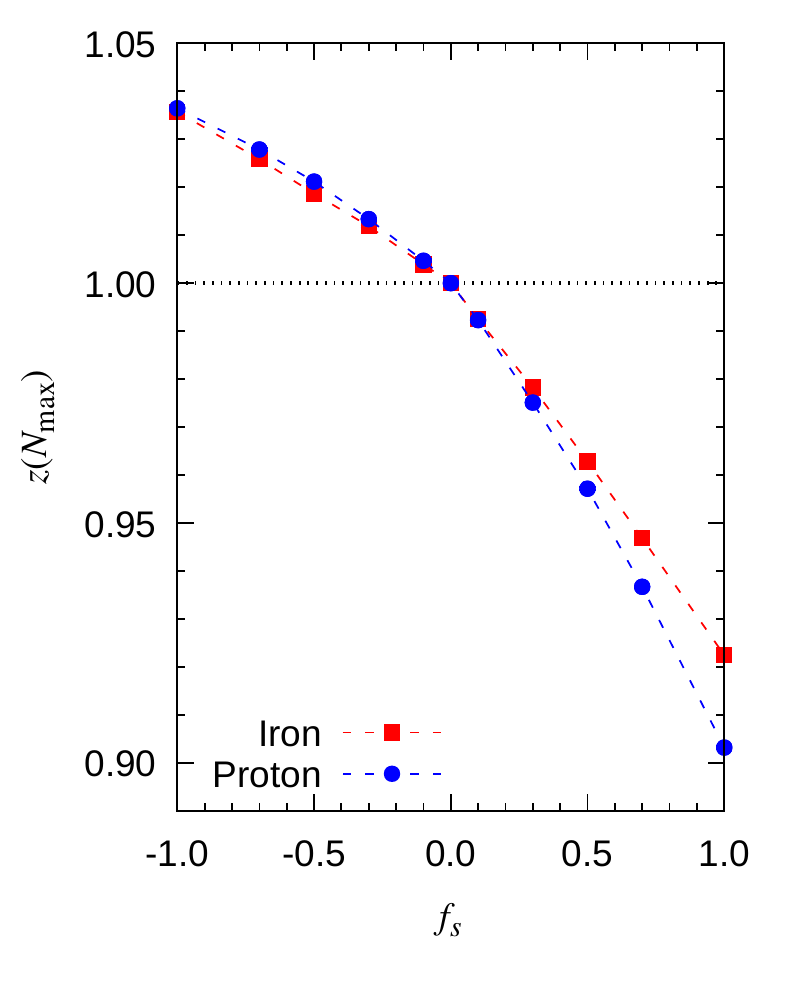}&
    \kern1em &
    \includegraphics[width=0.30\textwidth]{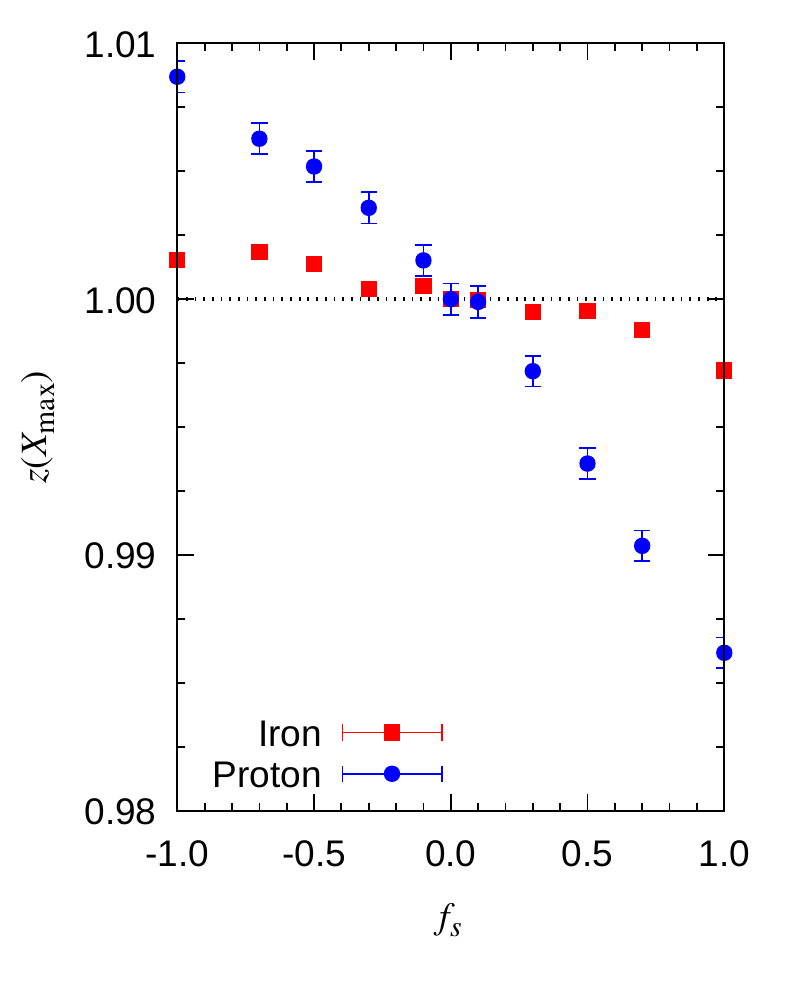}
\\*[-12pt]
\end{tabular}
\caption{$z(R_\mu)$, $z(N_{\rm max})$, and $z(X_{\rm max})$ as a function of
$f_s$, for $E_{\rm proj} = 10~{\rm EeV}$, $E_{\rm smin}=1~{\rm TeV}$,
and $E_{\rm pmin} = 1~{\rm PeV}$.  We have run 1600 (20000) showers
per point for ground muons (longitudinal development), setting at each
case the thinning algorithm parameters to get a more detailed
simulation of the hadronic or the electromagnetic cascade,
respectively. \label{fig:dos}}
\end{figure*}

During shower simulation, hadronic collisions are processed via calls
to an event generator; we adopt SIBYLL 2.3d~\cite{Riehn:2019jet}. The input parameters for these
calls are the projectile identity $p_{\rm id}$, its kinetic energy
$E_{\rm proj}$, and the target
identity. On return, the generator provides a list of $N_{\rm sec}$
particles, specifying their identity $s_{{\rm id}_i}$, energy $E_{{\rm sec}_i}$, momentum, etcetera, with $i=1,\cdots,N_{\rm sec}$.

\begin{figure*}
\centering
\begin{tabular}{ccccc}
  \includegraphics[width=0.30\textwidth]{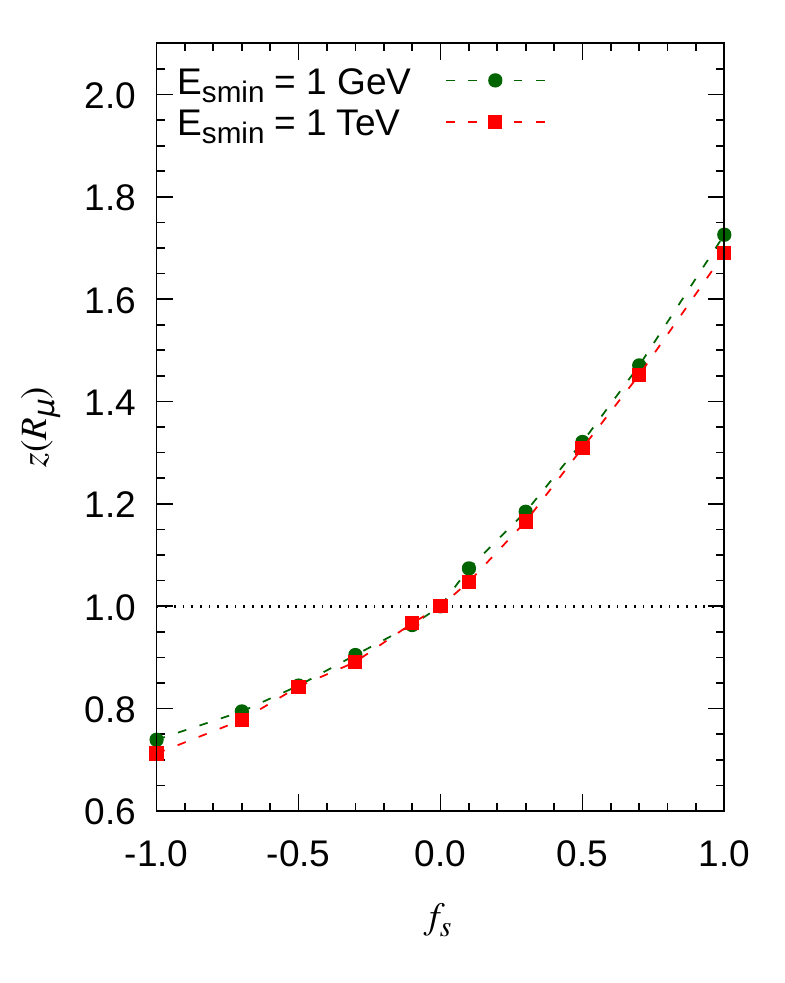} &
  \kern1em &
  \includegraphics[width=0.30\textwidth]{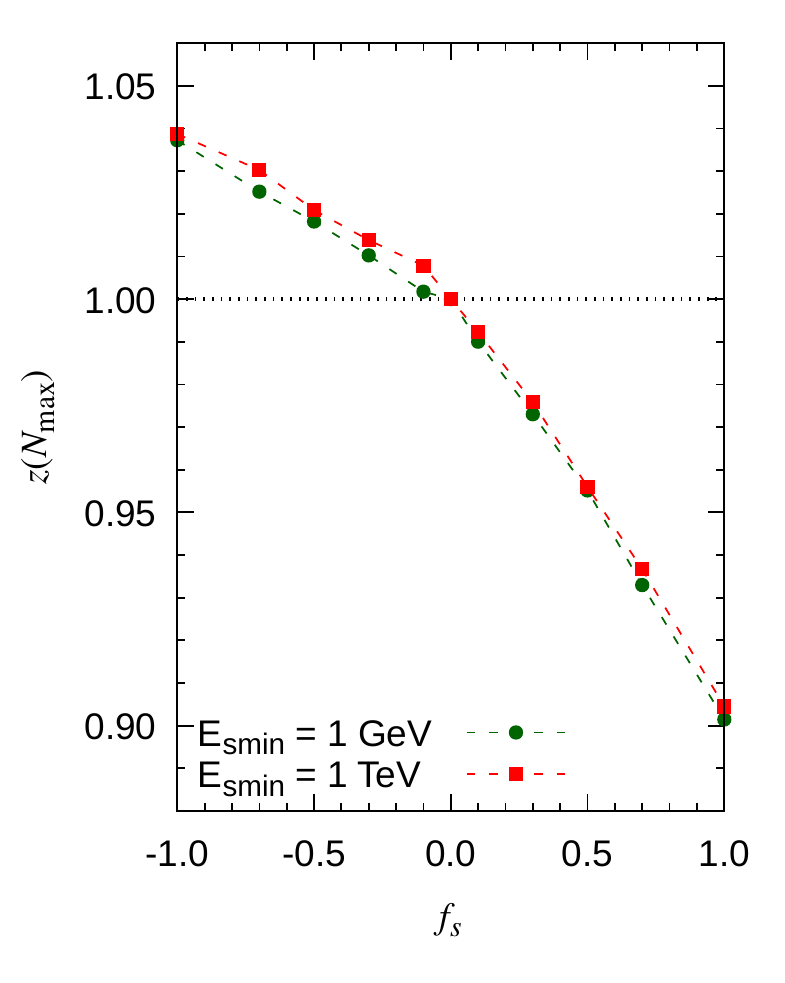} &
  \kern1em &
  \includegraphics[width=0.30\textwidth]{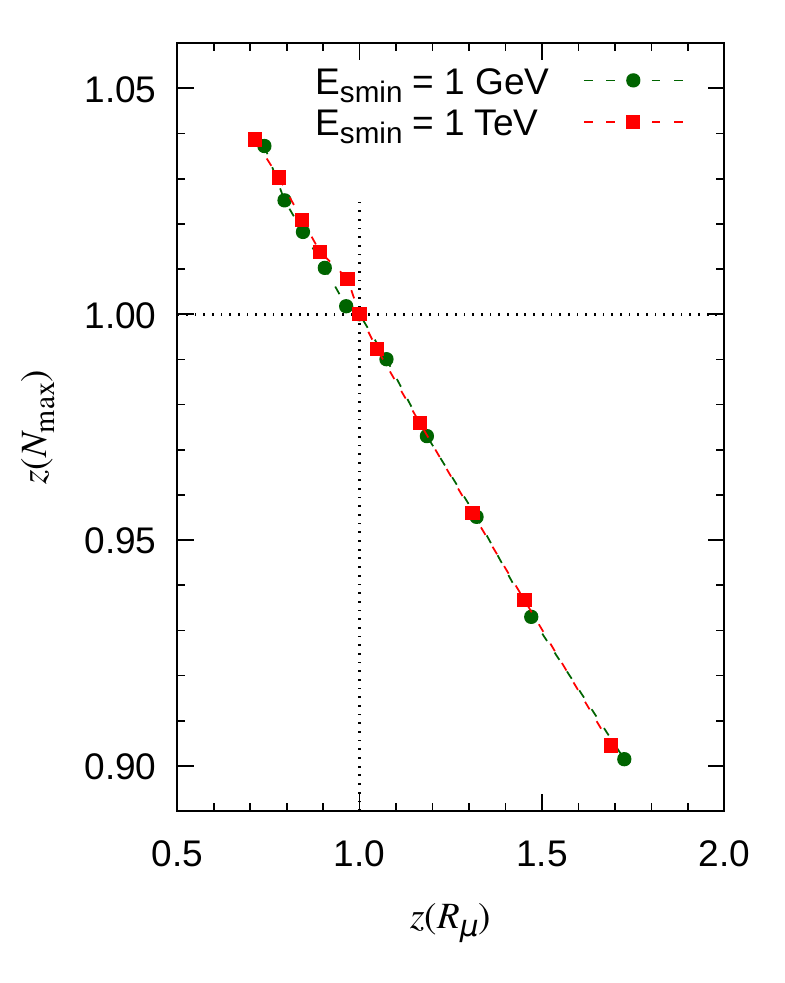}
\\
  \includegraphics[width=0.30\textwidth]{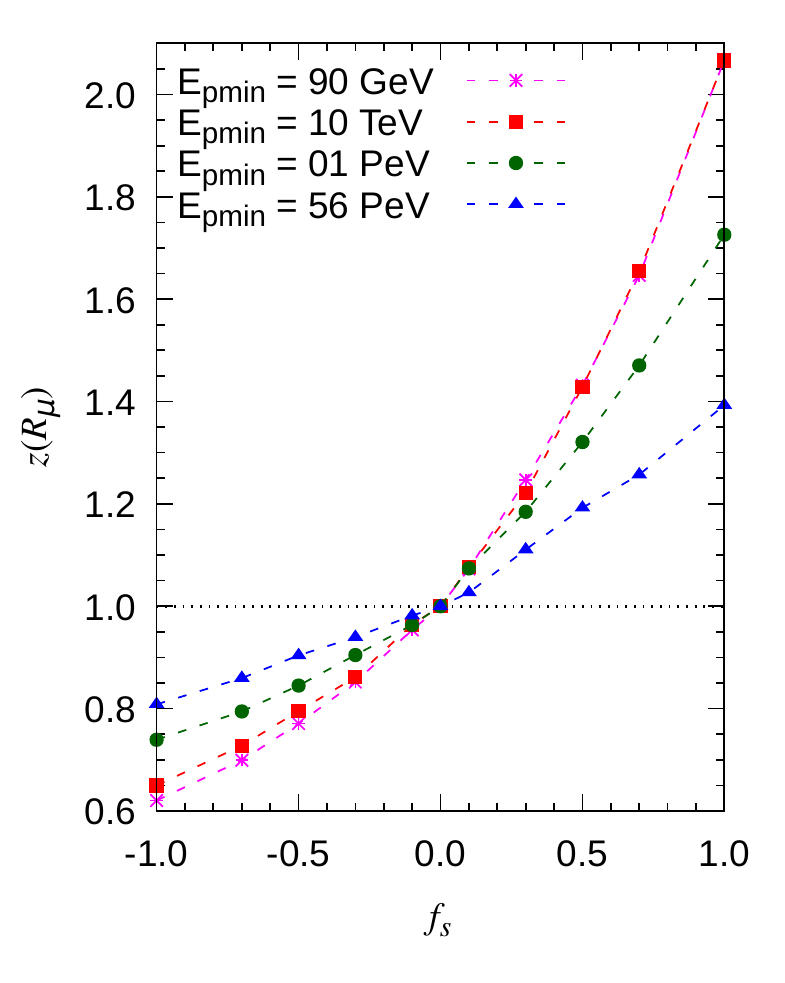} &
  \kern1em &
  \includegraphics[width=0.30\textwidth]{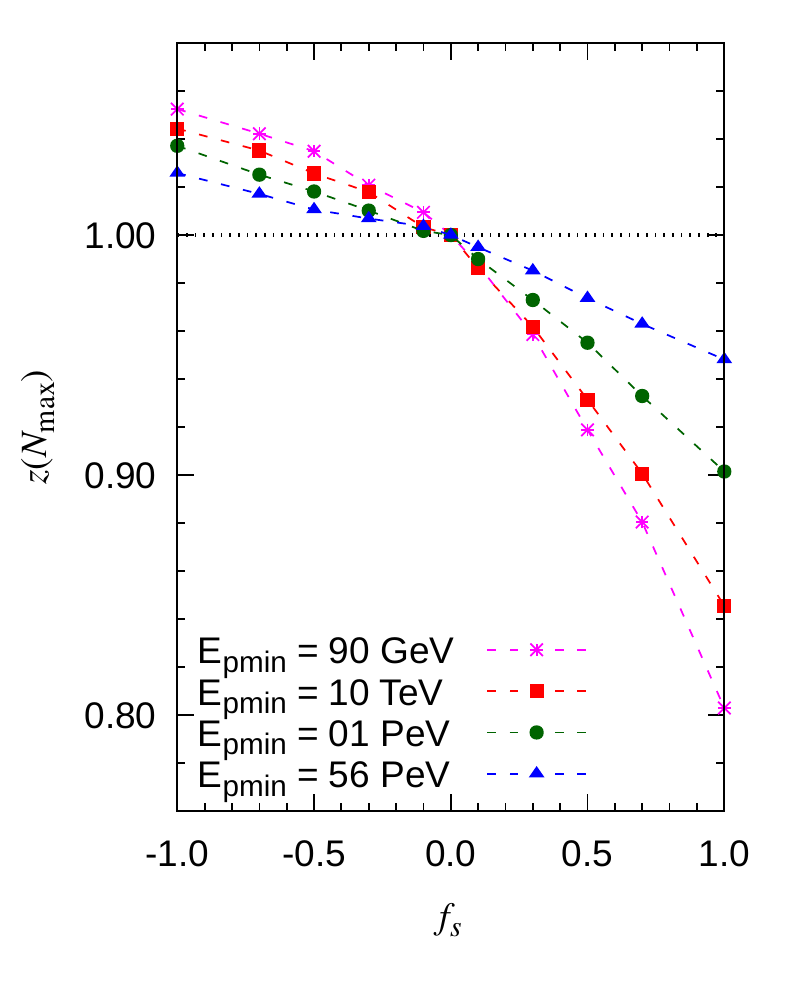} &
  \kern1em &
  \includegraphics[width=0.30\textwidth]{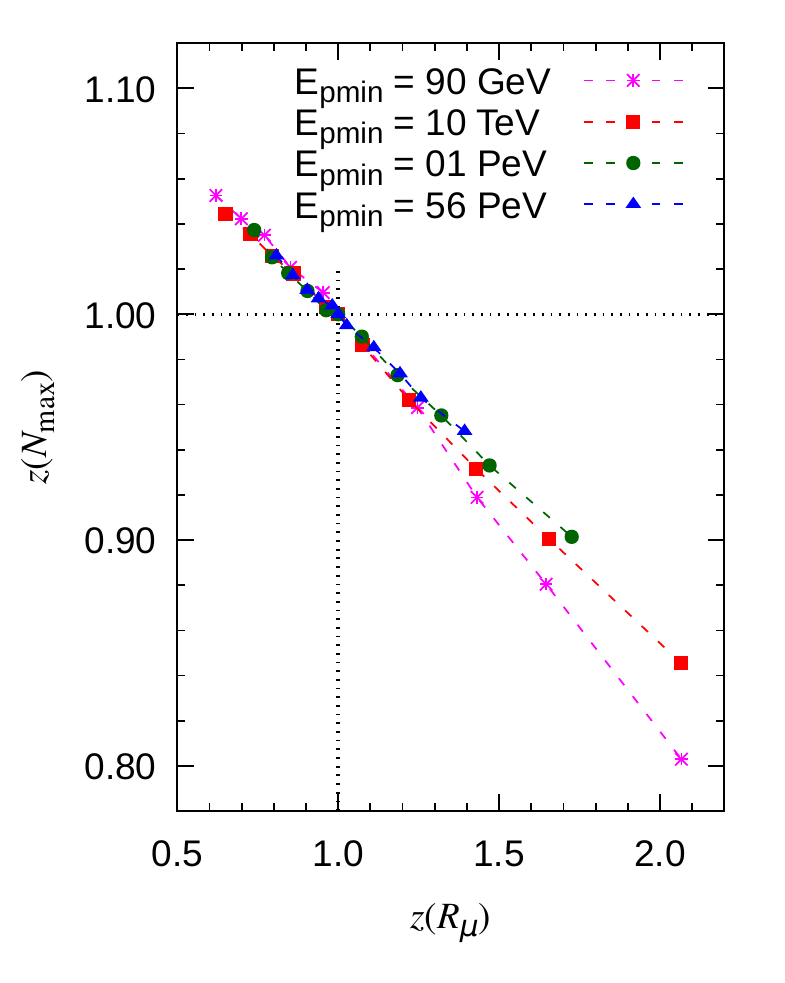}
\\*[-12pt]
\end{tabular}
\caption{$z(R_\mu)$ versus $f_s$ (left), $z(N_{\rm max})$ versus $f_s$
  (middle), and $z(R_\mu)$ versus  $z(N_{\rm max})$ (right), for varying $E_{\rm
    smin}$ (upper), and $E_{\rm pmin}$ (lower). \label{fig:trescuatro}}
\end{figure*}
All the returned secondary particle lists undergo a post-processing
process, just before they are stacked into the particle stacks for
further propagation. The post-processing algorithm obeys the following
rules:
\begin{enumerate}
\item If $f_s=0$ or $E_{\rm proj}<E_{\rm pmin}$ or $E_{\rm proj}>E_{\rm
  pmax}$ then no action is taken; the secondary particle list remains
  unchanged.
\item If $f_s\ne 0$ and $E_{\rm pmin} \le E_{\rm
  proj} \le E_{\rm pmax}$ then the list of secondaries is scanned,
  and processed as follows:
\begin{enumerate}
\item If $f_s > 0$, all the secondary pions whose kinetic energies lie
  within the interval $[E_{\rm smin},E_{\rm smax}]$ are considered for
  identity swapping. Each of them is randomly selected with
  probability $|f_s|$. In case of positive selection, the identity is
  changed with the following criteria:
\begin{enumerate}
\item Each $\pi^0$ is transformed onto $K^0_S$ of $K^0_L$, with 50\% chance between them.
\item Each $\pi^+$ ($\pi^-$) is transformed onto $K^+$ ($K^-$).
\end{enumerate}
\item If $f_s < 0$, all the secondary kaons whose kinetic energies lie
  within the interval $[E_{\rm smin},E_{\rm smax}]$ are considered for
  identity swapping. Each of them is randomly selected with
  probability $|f_s|$. In case of positive selection, the identity is
  changed with the following criterion:
\begin{enumerate}
\item Each $K^0_S$ or $K^0_L$  is transformed onto $\pi^0$.
\item Each $K^+$ ($K^-$) is transformed onto $\pi^+$ ($\pi^-$).
\end{enumerate}
\end{enumerate}
\item The kinetic energy of swapped particles is
 set so as to keep total energy conserved.
\end{enumerate}

\subsection{Air Shower Simulations}

\begin{figure*}[tpb]
\centering
\includegraphics[width=0.45\textwidth]{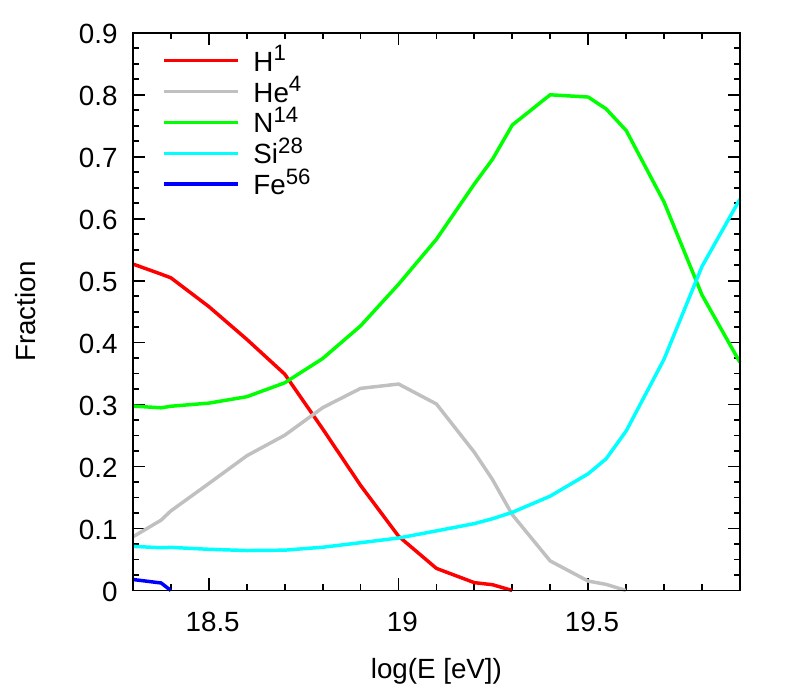}
\includegraphics[width=0.45\textwidth, trim={0 -1.5mm 0 4mm},clip]{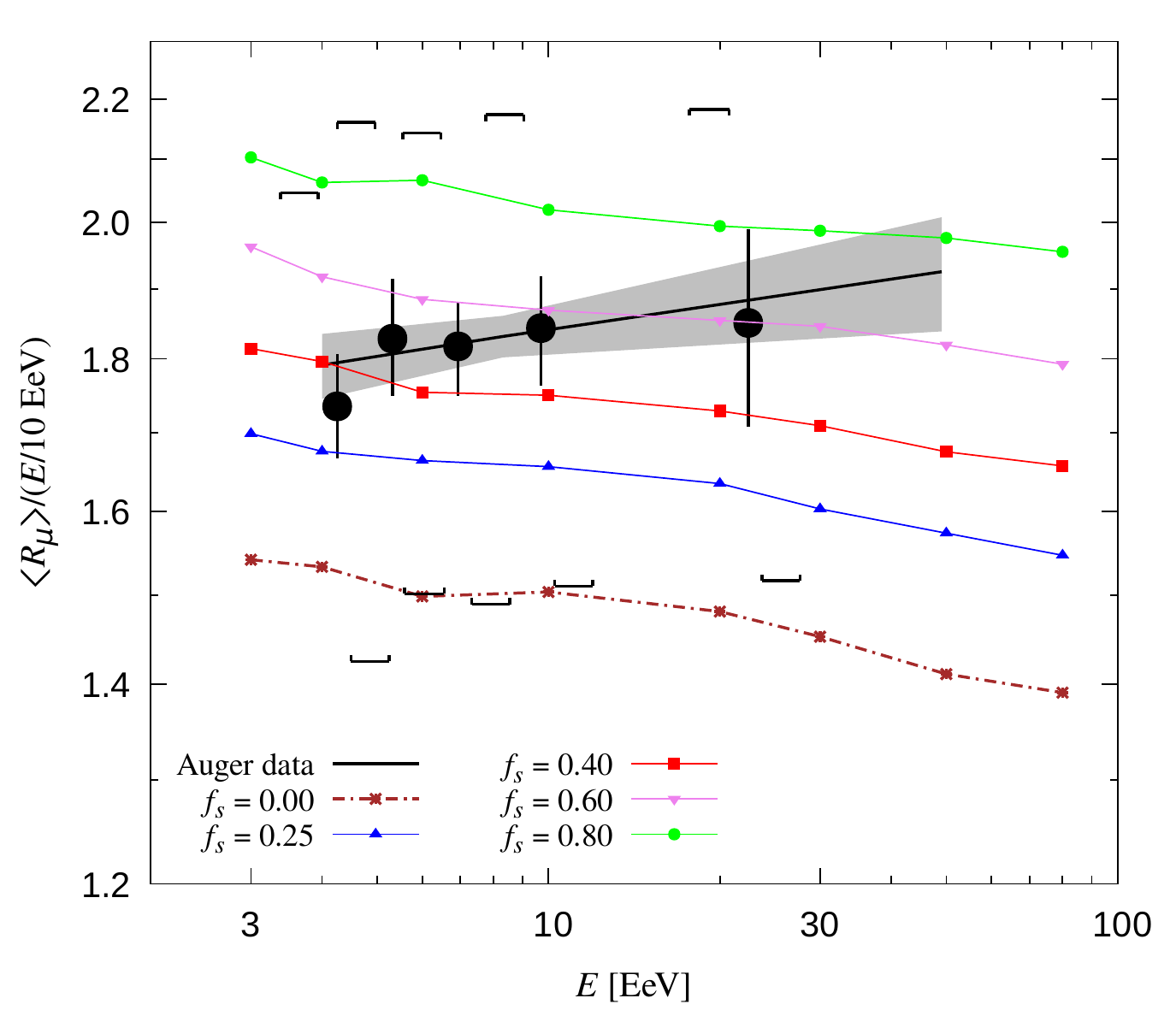}
\caption{{\bf Left.} Fractions of ultra-high energy primary cosmic
  rays entering at the top of the Earth’s atmosphere, as functions of
  the primary energy, evaluated from partial fluxes corresponding to
  the fit reported by the Pierre Auger Collaboration~\cite{PierreAuger:2016use}. {\bf
    Right.}~Estimations of $R_\mu$ from AIRES + SIBYLL 2.3d simulations for different values of
  $f_s$ superimposed over Auger data with statistical
  \mbox{($\hspace{0.1em}\bullet\hspace*{-.66em}\mid\hspace{0.16em}$)}
  and systematic (\protect\rotatebox{90}{\hspace{-.075cm}[ ]})
  uncertainties~\cite{Aab:2014pza}. We have adopted the mixed
  baryonic composition shown in the left panel. \label{fig:cinco}}
\end{figure*}

To characterize the possible cross-correlation among selected
observables we have simulated more than a million showers with incident zenith angles of $45^\circ$ and $67^\circ$. The
shower observables relevant to our analysis are:
\begin{itemize}[noitemsep,topsep=0pt]
\item the depth of maximum shower development $X_{\rm max}$ and its
ﬂuctuations $\sigma X_{\max}$;
\item the dimensionless muon content $R_\mu = N_\mu/N_{\mu, 19}$ and its fluctuations $\sigma R_\mu$, where $N_\mu$ is the total number of muons (with $E_\mu >300~{\rm MeV}$) at
  ground level and $N_{\mu,19} =  1.455 \times 10^7$
  is the average number of muons in simulated proton showers at
  $10^{19}~{\rm eV}$ with incident angle of $67^\circ$; 
\item the number of charged particles at the shower maximum $N_{\rm
    max}$. 
\end{itemize}
For each observable  ${\cal O}$, we evaluate  
\begin{equation}
z({\cal O}) = \frac{\langle{\cal O}(f_s)\rangle}{\langle{\cal O}(f_s=0)\rangle}
\, , 
\end{equation}
to work with normalized variables.

In Fig.~\ref{fig:dos} we show $z(R_\mu)$, $z(N_{\rm max})$, and $z(X_{\rm
max})$, as a function of $f_s$, for
$E = 10~{\rm EeV}$, $E_{\rm smin}=1~{\rm TeV}$, and
$E_{\rm pmin} = 1~{\rm PeV}$, with both $E_{\rm smax}$ and $E_{\rm pmax}$ set to infinite. Note that this particular $E_{\rm pmin}$  corresponds to
hadronic interactions at $\sqrt{s_{NN}} \approx 1.41~{\rm TeV}$, just
below the energy ($\left. \sqrt{s_{NN}} \right|_{_{\rm ALICE}} \simeq 2.76~{\rm TeV}$)
 where the ALICE Collaboration reported a smooth rise of the hyperon-to-pion ratio~\cite{ALICE:2013xmt}. Note also
that for $f_s<0$, kaons are changed into
pions, whereas for $f_s>0$, pions are changed into kaons, with
progressive probability equal to $|f_s|$. The simulations to evaluate $X_{\rm
    max}$ are always carried out using inclined showers at $45^\circ$.
The variations in $X_{\rm max}$ fluctuations (not shown in the figure) are very small: $|z(\sigma X_{\rm max}) -1|<0.03$ for all
$f_s\in [-1,1]$.
 Taking $f_s \sim 0.4$ as fiducial we observe a
change in $R_\mu$ of roughly 20\% for showers initiated by protons
and 10\% in those initiated by iron. These variations correspond to a
reduction of  $N_{\rm max}$ by about 3\%. In the right panel of
Fig.~\ref{fig:dos} we can see that the model predictions on $X_{\rm
  max}$ vary less than 1.5\% when compared to the $f_s=0$ result.
 Similarly, the fluctuations $\sigma X_{\rm max}$ vary
 by less than 3\%.
 Our analysis thus corroborates the results presented in~\cite{Allen:2013hfa}, which show
that by suppressing the $\pi^0$ energy fraction we can obtain an increase in the number of muons
at ground without coming into conflict with $X_{\rm max}$ observations.

To study the model dependence with $E_{\rm smin}$ and $E_{\rm pmin}$
we use proton induced showers.  In Fig.~\ref{fig:trescuatro} we show the
dependences of $R_\mu$ and $N_{\rm max}$ with $E_{\rm smin}$ (upper row) and $E_{\rm pmin}$
(lower row). We can see that the change of
$E_{\rm smin}$ leads to negligible effects, and that there is virtually no
difference between $E_{\rm pmin} = 90~{\rm GeV}$ and
$E_{\rm pmin} = 10~{\rm TeV}$, indicating a saturation effect; see
Appendix~\ref{appA}. These are, however, unrealistic energy thresholds.  A linear
dependence between the two observables is evident, especially for
$z(R_\mu) \sim 1$. The physically unrealistic case of
$E_{\rm pmin}=90~{\rm GeV}$ is the one that presents the largest
departure from linearity.

\begin{figure}[tpb]
\centering
\includegraphics[width=0.45\textwidth]{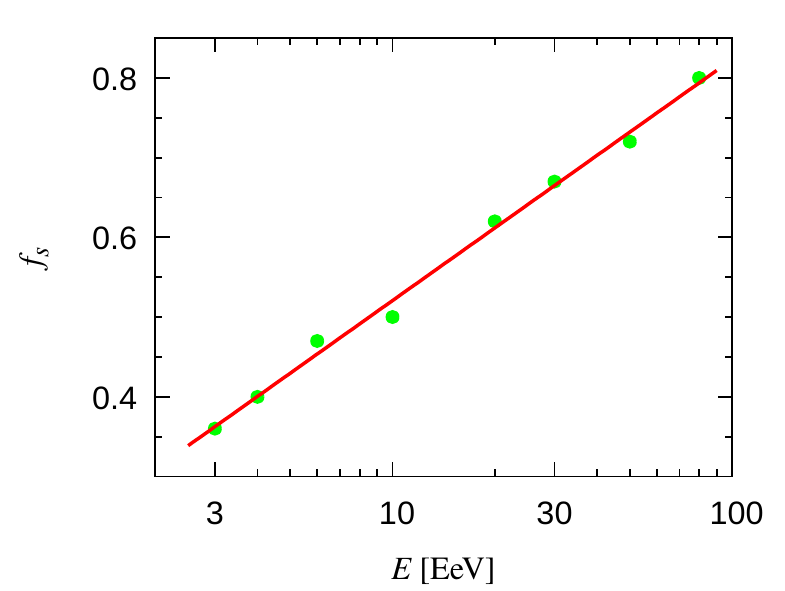}
\caption{Variation of $f_s$ with energy as determined from 
  eyeball fitting the rising straight-line $R_\mu$ estimation of
  Fig.~\ref{fig:cinco}. \label{fig:seis}}
\end{figure}

In the spirit of~\cite{Sciutto:2019pqs}, we now incorporate  the
change of the nuclear composition of the cosmic ray primary~\cite{PierreAuger:2016use} and study the variation of
$\langle R_\mu \rangle/ (E/10~{\rm EeV})$. As displayed in
Fig.~\ref{fig:cinco}, the effect of increasing $R_\mu$ yields a
flattening of the curve when compared to the   $f_s=0$
prediction. Even though $f_s \sim 0.4$ seems to roughly accommodate
the data around $E \sim
10^{19}~{\rm eV}$, it is clear from the shape of the best-fit curve
that to describe the muon anomaly in a larger energy range we would
need an energy-dependent $f_s$; see Fig.~\ref{fig:seis}.

We note, however, that this zeroth order approximation should be
understood as an effective (macroscopic) description of the entire
shower evolution, rather than a collection of individual interactions
generated by a homogeneous beam of projectiles. In this approach
$E_{\rm pmin}$ is no less important than $E_{\rm smin}$ and for a
$10^{10}~{\rm GeV}$ proton shower with $f_s = 0.7$ the number of pions
effectively swapped barely exceeds 0.5\% of the total number of
secondaries generated in shower. Global observables, such as the
number of muons at ground level, were obtained after adding and
averaging heaps of individual contributions, a process in which
statistics erases many ``microscopic'' details.

\begin{figure*}
\centering
\begin{tabular}{ccc}
\includegraphics[width=0.4\textwidth]{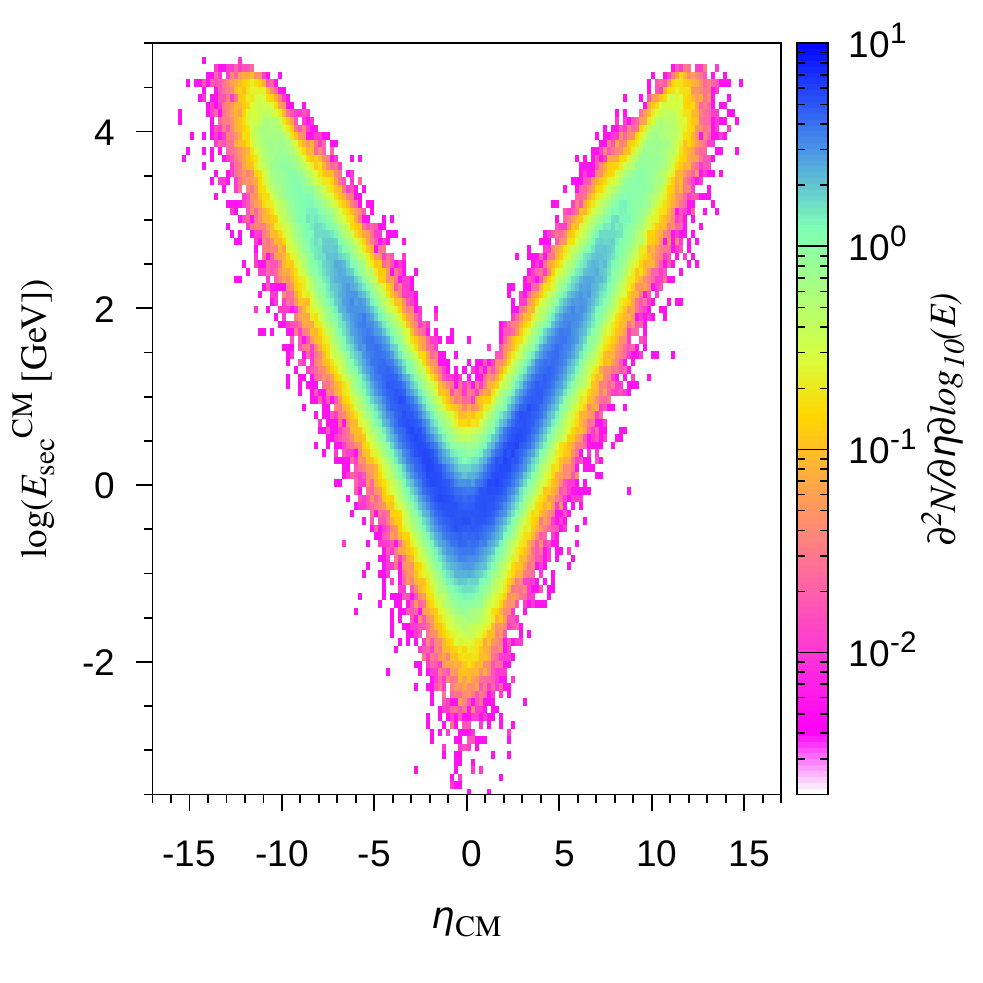}&
\kern3em &
\includegraphics[width=0.4\textwidth]{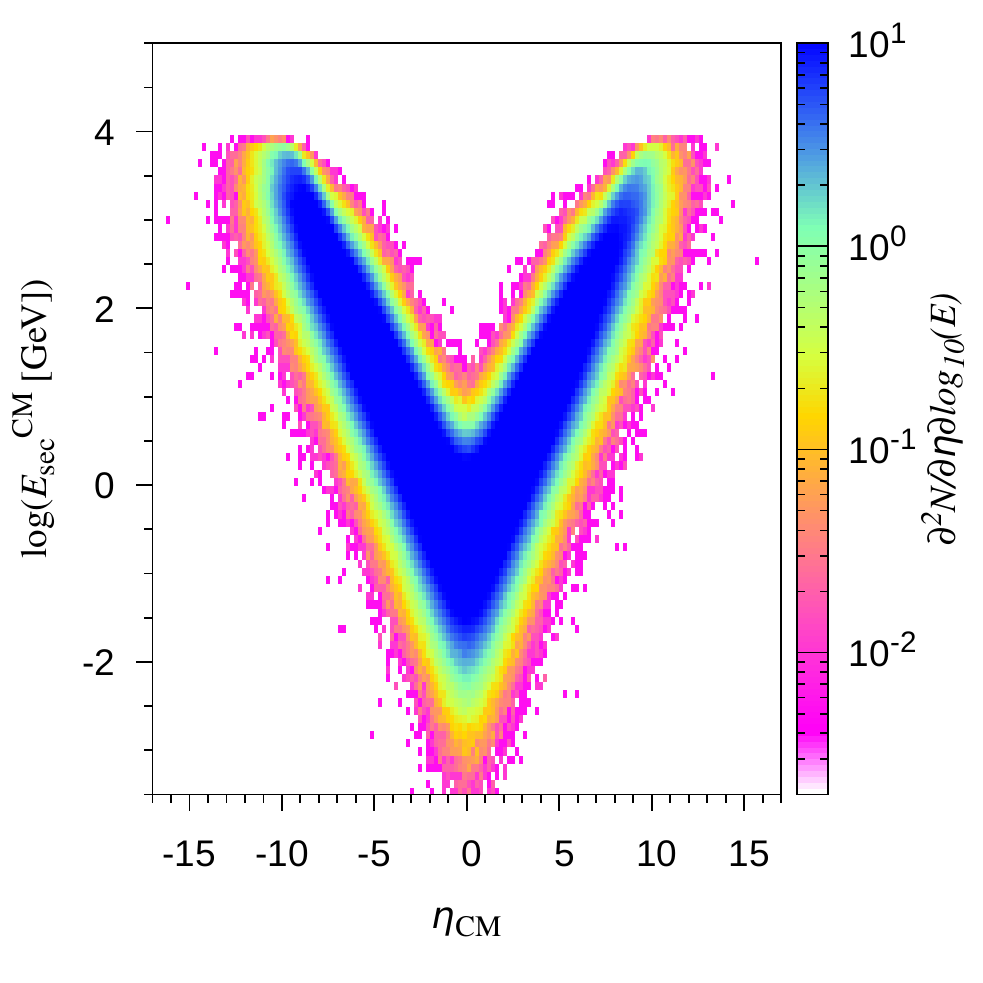}
\\
\includegraphics[width=0.4\textwidth]{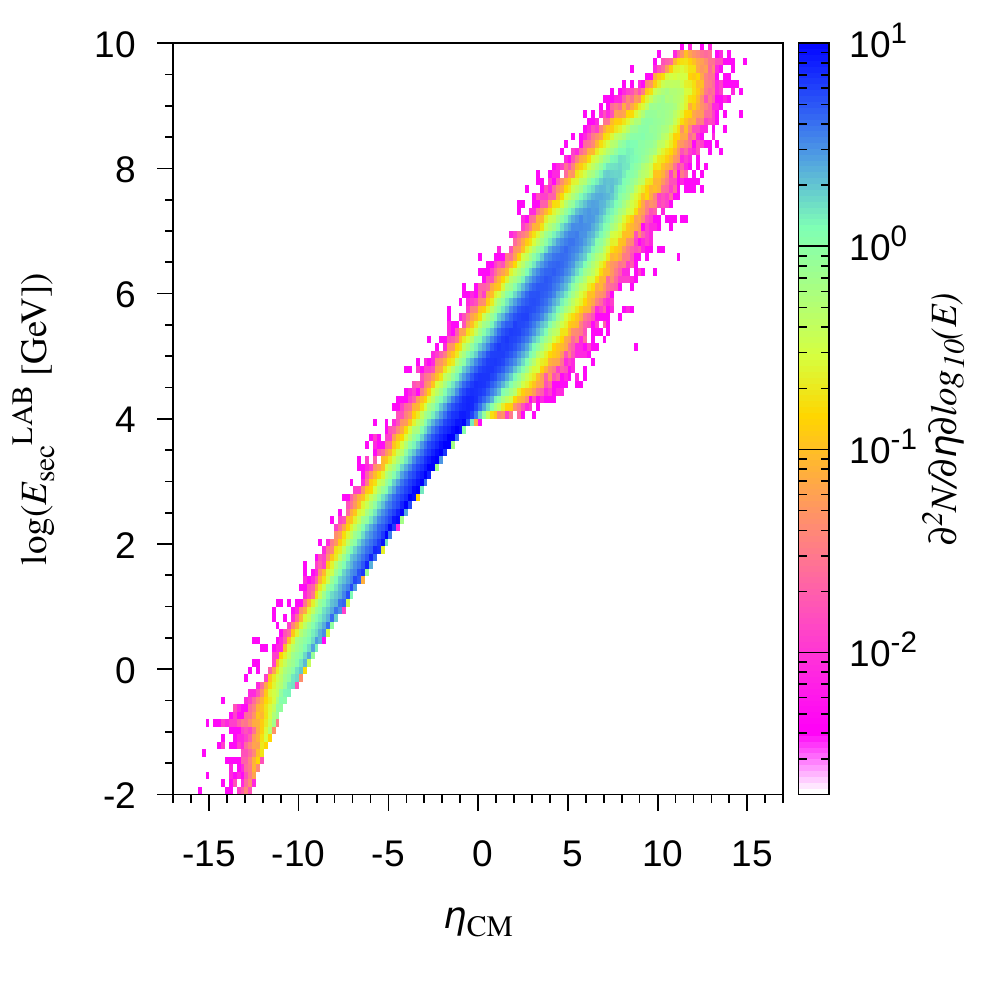}&
\kern3em &
\includegraphics[width=0.4\textwidth]{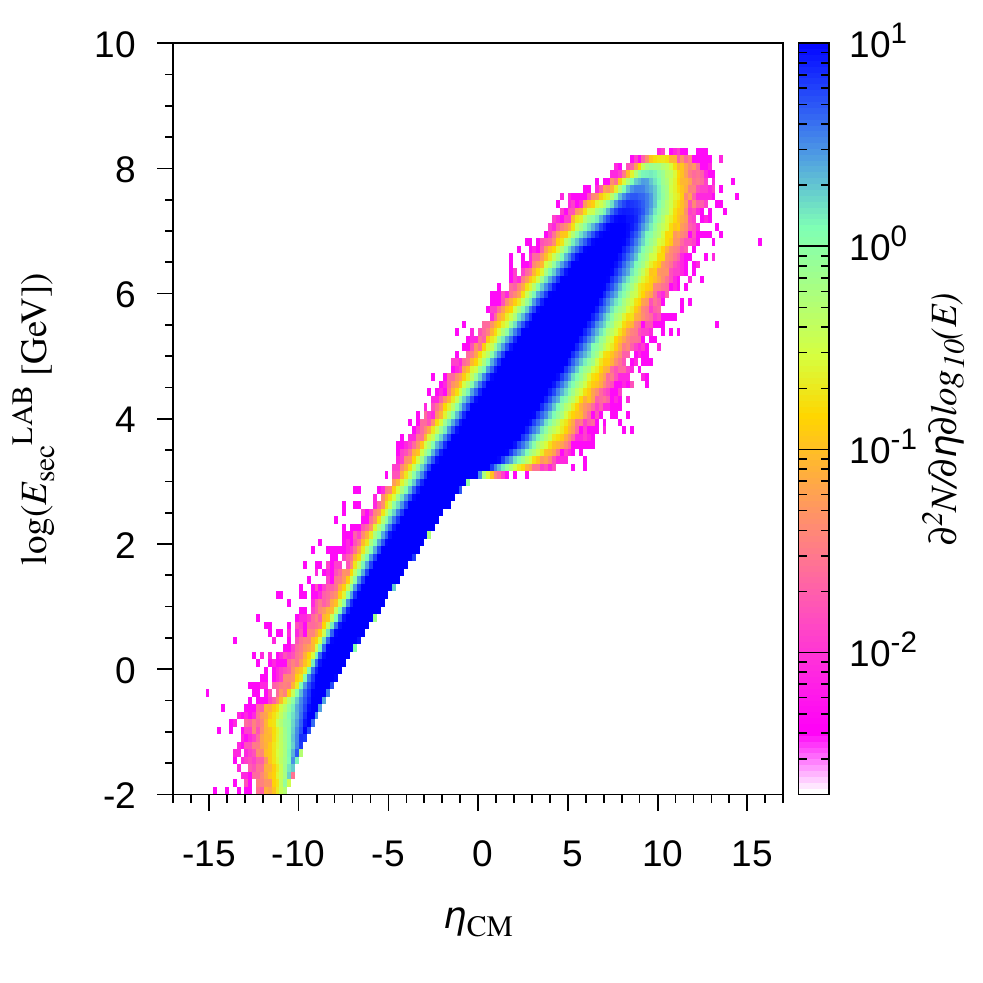}
\end{tabular}
\caption{Pion $E_{\rm sec}^{\rm CM}$ vs $\eta_{\rm CM}$ (upper row)
and $E_{\rm sec}^{\rm LAB}$ vs $\eta_{\rm CM}$ (lower row) bivariate
  distributions. The left (right) column correspond to the results from $10^4$
  collisions of a 10 EeV proton (iron nucleus) scattering
  off a proton (nitrogen nucleus) at
  rest, simulated with SIBYLL 2.3d. \label{fig:3A12}}
\end{figure*}

\section{Model Refinement}
\label{sec:3}

In the previous section we have shown that the zeroth order
approximation toy model gives a fair description of all shower
observables. However, there are two important caveats with this toy
model. Firstly, heavy flavor production should be enhanced in
kinematic regimes where quark masses may be insignificant. This
implies that a more realistic parametrization of $F_s$, which can
accurately describe single particle collisions, should depend on
pseudorapidity. Secondly, the shape of the best-fit curve to
Auger data is driven by both strangeness enhancement and the rapid
change in the nuclear composition~\cite{Sciutto:2019pqs}. Thus,
nuclear effects~\cite{Anchordoqui:2016oxy} could play a conclusive
role in bridging the gap between data and simulations, hinting that
$F_s$ should also have a variation with the nucleus baryon number
$A$. Along this line, a strong suppression of the production of
neutral pions in $p$Pb collisions was reported by the LHCf
Collaboration after comparing to the results of $pp$
scattering~\cite{LHCf:2014gqm}. Uncertainties on the $A$ dependence of
$F_s$ are still quite large, and so for simplicity, we will neglect
$A$-induced effects in our study. Future LHC data (including $p$O and
OO collisions~\cite{Citron:2018lsq}) will provide new insights to
reduce these uncertainties and guide software
development.
\begin{figure*}
\begin{center}
\begin{tabular}{ccc}
\includegraphics[width=0.4\textwidth]{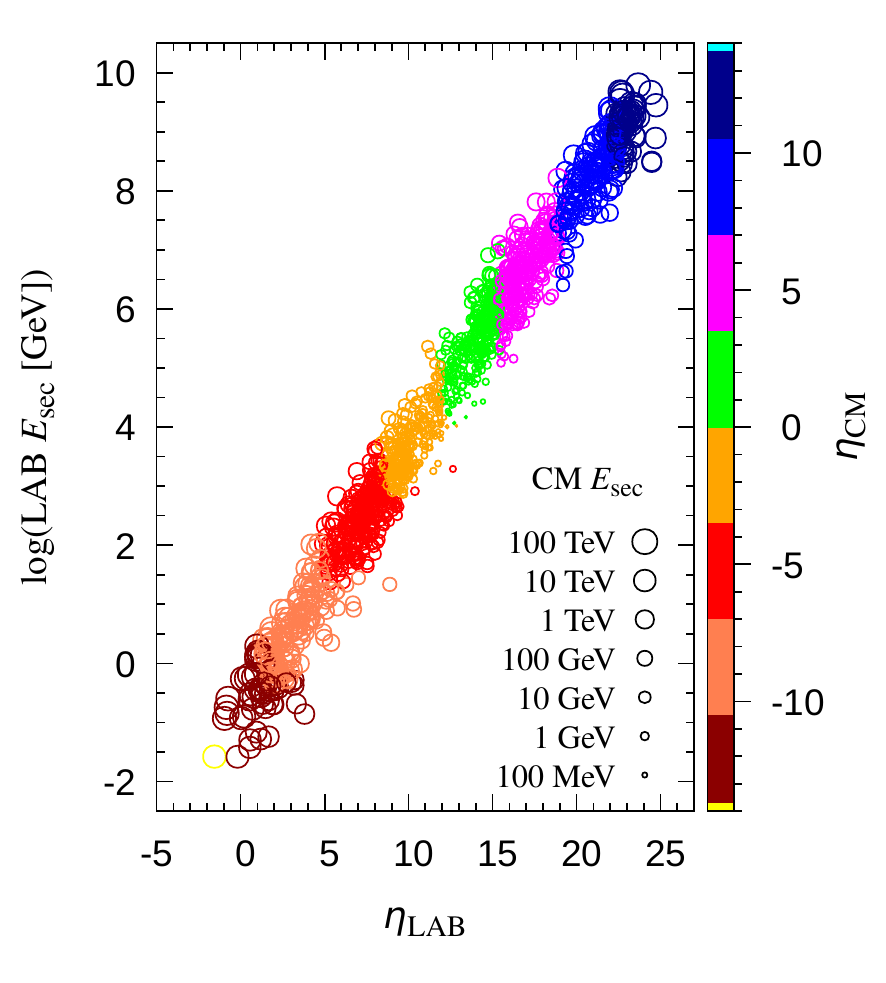}&
\kern2em &
\includegraphics[width=0.4\textwidth]{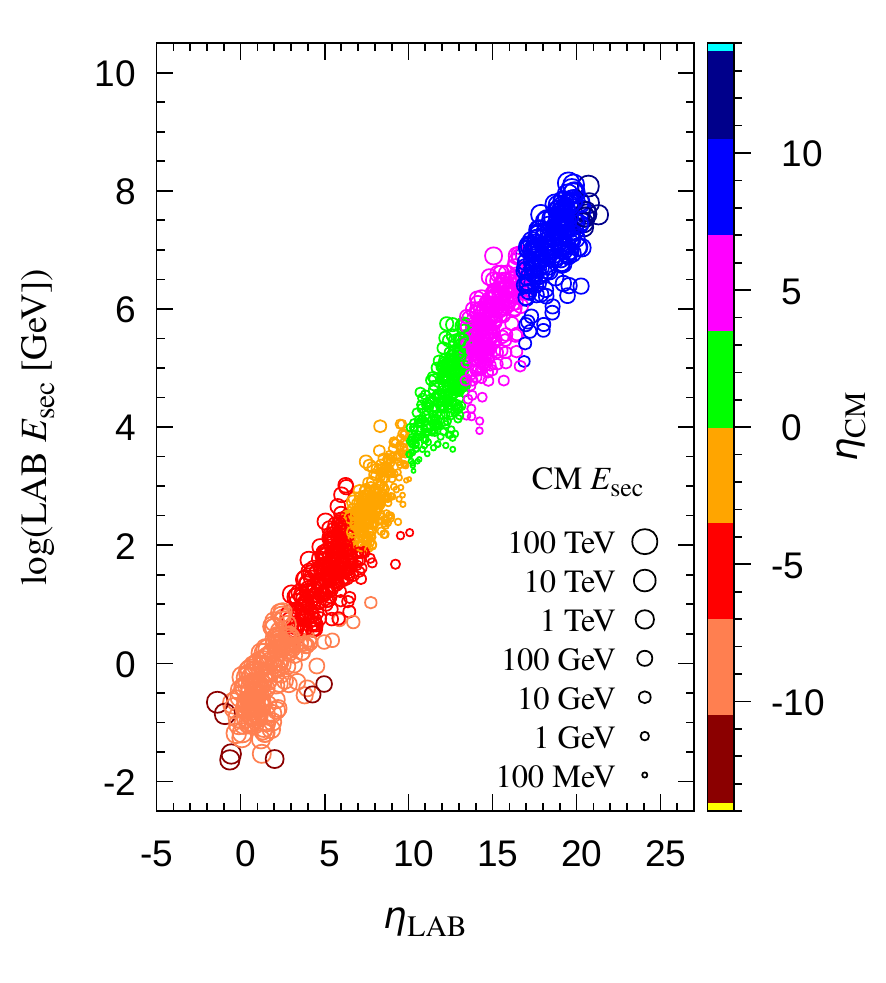}
\\*[-15pt]
\end{tabular}
\end{center}
\caption{$E_{\rm sec}^{\rm LAB}$ vs $\eta_{\rm LAB}$ scatter plots for
  secondary pions generated in 10 EeV collisions of a proton
  scattering off a stationary proton (left) and an iron nucleus scattering of
  a nitrogen nucleus (right). The CM pseudorapidity and CM kinetic energy of the secondaries
  can be appreciated at each plotted dot by means of the dot color and its
  size, respectively, accordingly with the color scale placed at
  the right of the plots and size scales indicated in each graph. \label{fig:3A3}}
\end{figure*}

The Lorentz transformation between the
center-of-mass (CM) and laboratory (LAB) systems is given by
\begin{equation}
  E_{\rm LAB} = \gamma (E_{\rm CM} + \beta \ p_{\rm long, CM}) \,,
\end{equation}
where $\gamma$ is the Lorentz factor and $\beta$ the velocity of the
CM with respect to the LAB frame. For ultrarelativistic
particles, $\beta \sim 1$ and
$p_{\rm long, CM} \sim E_{\rm CM} \ \cos \theta_{\rm CM}$, where
$\theta_{\rm CM}$ is
the angle of the secondary particle's momentum with respect to the
axis where the projectile of the collision moves (i.e. direction of
the beam). A straightforward substitution leads to
\begin{equation}
  E_{\rm LAB} \sim \gamma \ E_{\rm CM} \  (1 + \cos \theta_{\rm CM})  \, .
\label{ElabEcm}
\end{equation}
At first sight one may conjecture that the imposed lower limit on $E_{\rm smin}$ in our toy
model is inconsistent with the description of hadronic 
collisions as  $0< E_{\rm LAB}< 2\gamma E_{\rm CM}$. To inspect the forward-backward directions in the CM frame we conveniently work with the pseudorapidity
\begin{equation}
  \eta_{\rm CM} = - \ln \left[\tan \left(\frac{\theta_{\rm CM}}{2} \right) \right] \, .
\end{equation}
The forward-backward symmetry of Eq.~(\ref{ElabEcm})  is evident in
the pion pseudorapidity
distributions shown in the upper row of Fig.~\ref{fig:3A12}
  We note that the toy model approximation $E_{\rm
  smin} = 1~{\rm TeV}$ breaks this symmetry when going into
  the LAB frame; see the lower row of
Fig.~\ref{fig:3A12}. In particular, pions with $\eta_{\rm CM} < -4$ are not
considered for swapping in the AIRES module described in
Sec.~\ref{sec:2}. The relation between the CM and LAB pseudorapidity
is displayed in the scatter plots of Fig.~\ref{fig:3A3}. It is
important to stressed that the densities of dots in different places
  of these plots may not accurately represent the
    actual number of secondaries that corresponds to each location
    within the $(\eta_{\rm LAB},E_{\rm LAB})$ plane. This is due to
  the fact that to improve the graphics readability, only a small
  fraction, non-uniformly sampled, of the total number of secondaries
  produced in the collision has been represented. The sampling was
  performed trying to obtain a uniform coverage of the entire range of
  CM pseudorapidities of the secondaries. To this end, the $-\infty
  <\eta_{\rm CM}<\infty$ axis is partitioned in consecutive intervals,
  with extremes at the points $-\infty, -10, -7, -5, -4, -3, -2,
  0, 2, 3, 4, 5, 7, 10, \infty$, and then the entire set of secondary pions emerging
  from the collisions is scanned sampling 100 cases for each one of
  those intervals.  For a realistic appreciation of the distribution
  of secondary particles, it is better see the bivariate distributions shown in
  Fig.~\ref{fig:3A12}.

  As the shower develops in the atmosphere, the hadrons propagate
  through a medium with an increasing density while the altitude
  decreases and the hadron-air cross section rises slowly with
  energy. Thereby, the probability for interacting with the air
  molecules before decay increases with rising energy. Furthermore,
  the relativistic time dilation increases the decay length by a
  factor $E_h/m_h$, where $E_h$ and $m_h$ are the energy and mass of
  the produced hadron.  The $\pi^0$'s, with a lifetime of
  $\simeq 8.4 \times10^{-17}~{\rm s}$, do decay promptly to two
  photons, feeding the electromagnetic component of the shower. To see
  how neutral kaons could suppressed this process, it is instructive
  to estimate the critical energy at which the chances for interaction
  and decay are equal for other longer-lived mesons. For a vertical
  transversal of the atmosphere, the critical energy is found to be:
  $\xi_c^{\pi^\pm} \sim 115~{\rm GeV}$,
  $\xi_c^{K^\pm} \sim 850~{\rm GeV}$,
  $\xi_c^{K^0_L} \sim 210~{\rm GeV}$,
  $\xi_c^{K^0_S} \sim 30~{\rm TeV}$~\cite{Gondolo:1995fq}.  The
  dominant $K^+$ branching ratios are to $\mu^+ \nu_\mu\ (64\%)$, to
  $\pi^+ \pi^0\ (21\%)$, to $\pi^+\pi^+\pi^-\ (6\%)$, and to
  $\pi^+ \pi^0\pi^0\ (2\%)$, whereas those of the $K^0_S$ are to
  $\pi^+\pi^- \ (60\%)$, to $\pi^0 \pi^0 \ (30\%)$, and for $K^0_L$ we
  have $\pi^\pm e^\mp \nu_e \ (40\%)$,
  $\pi^\pm \mu^\mp \nu_\mu \ (27\%)$, $\pi^0 \pi^0 \pi^0 \ (19\%)$,
  $\pi^+ \pi^- \pi^0 \ (12\%)$~\cite{ParticleDataGroup:2020ssz}.
  Using these branching fractions, to a first approximation we can
  estimate that in each generation of particles about 25\% of the
  energy is transferred to the electromagnetic shower, and all hadrons
  with energy~$\agt\xi_c^{\pi^\pm}$ interact rather than decay,
  continuing to produce the hadronic shower~\cite{Anchordoqui:1998nq,Ulrich:2010rg}. Eventually, the
  electromagnetic cascade dissipates around 90\% of the primary
  particle's energy and the remaining 10\% is carried by muons and
  neutrinos. Even though these numbers depend on the incident zenith
  angle of the primary cosmic ray we note that very low energy kaons
  will decay before interacting to feed the electromagnetic showers in
  similar way neutral pions do. Therefore, the required symmetry with
  respect to the CM pseudorapidity seems to indicate that there {\it
    must} be swapping of some pions which do not produce an overall
  effect on the shower evolution. Taking these considerations into
  account, we are ready to amend the AIRES module.

Before proceeding, we pause to note that we have verified that there
is no significant difference in the scattering predictions by changing
the hadronic interaction model. For a direct comparison, in
Appendix~\ref{appB} we show the pion, kaon, and nucleon bivariate
distributions for the same particle collisions, but simulated with
EPOS-LHC~\cite{Pierog:2013ria}.

In what follows we refer to the measurements/tunes performed in the
``central'' and ``forward'' regions, as defined with respect to the CM
pseudorapidity of the particles. The central pseudorapidity region is
defined as $|\eta_{\rm CM}| \leq 2.5$, corresponding to the ATLAS~\cite{ATLAS:2008xda}, CMS~\cite{CMS:2008xjf}
and ALICE~\cite{ALICE:2008ngc} acceptances, and the forward pseudorapidity region as $|\eta_{\rm CM}|
 \geq 2.5$. It is generally thought that the observed differences
 between data and simulation originate, in most part, due to the model extrapolation from
the central pseudorapidity region, in which the hadronic event
generators adopted in UHECR shower simulations are mainly tuned. We
therefore assume herein that the enhancement of strangeness production is
negligible for $|\eta_{\rm CM}| <4$ (more on this below).  The free parameters of the refined model are defined as follows:
\begin{widetext}
\begin{tabular}{p{0.3\textwidth}cp{0.56\textwidth}}
{\bf Swapping probability} \dotfill      & $F_s(\eta_{\rm CM})$ &
     Controls the number of secondary pions that are
     affected by change of identity. $F_s$ depends on the centre of
     mass pseudorapidity of the secondary particles, $\eta_{\rm CM}$,
                                                         and must
                                                         verify $0\le
                                                         F_s \le
                                                         1$.
    Unless otherwise specified, we use
    \begin{equation}
       F_s(\eta_{\rm CM}) =\left\{
         \begin{array}{llc}
           f_s & \hbox{if} & -\infty < \eta_{\rm CM} < -4 \\*[3mm]
           0 & \hbox{if} & -4 \le \eta_{\rm CM} \le 4 \\*[3mm]
           f_s & \hbox{if} & \phantom{-!} 4 < \eta_{\rm CM} < \infty
           \end{array}
           \right. \,,
          \label{Fs}
    \end{equation} 
    with $0\le f_s \le 1$.         
\\*[10pt]
{\bf Minimum projectile energy} \dotfill & $E_{\rm pmin}$ &
     Particle swapping is performed in hadronic collisions whose
     projectile kinetic energy is larger than this energy. $E_{\rm
       pmin}$ must be larger than 900 MeV.
       As in our toy model we take $E_{\rm pmin} = 1~{\rm PeV}$.
\\*[10pt]
{\bf Minimum secondary energy} \dotfill & $E_{\rm smin}$ &
     Secondary particles with kinetic energies below this threshold
     are always left unchanged.  $E_{\rm smin}$ must be larger than 600~MeV. To sample the entire CM pseudorapidity region we take
                                                  $E_{\rm smin} =
                                                  1~{\rm GeV}$. 
\end{tabular}
\end{widetext}
The logics of the  hadronic collision post-processing remains the same
to that discussed in Sec.~\ref{sec:2b}.

\begin{figure*}
\centering
\begin{tabular}{ccc}
  \includegraphics[width=0.4\textwidth]{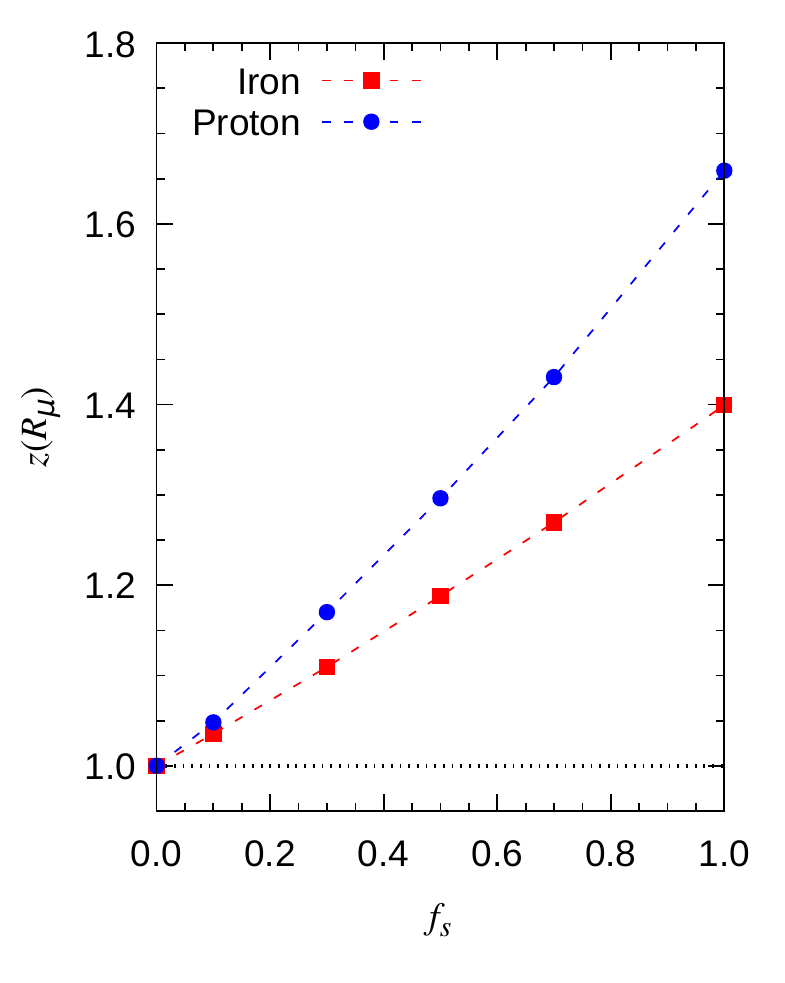}&
  \kern2em &
  \includegraphics[width=0.4\textwidth]{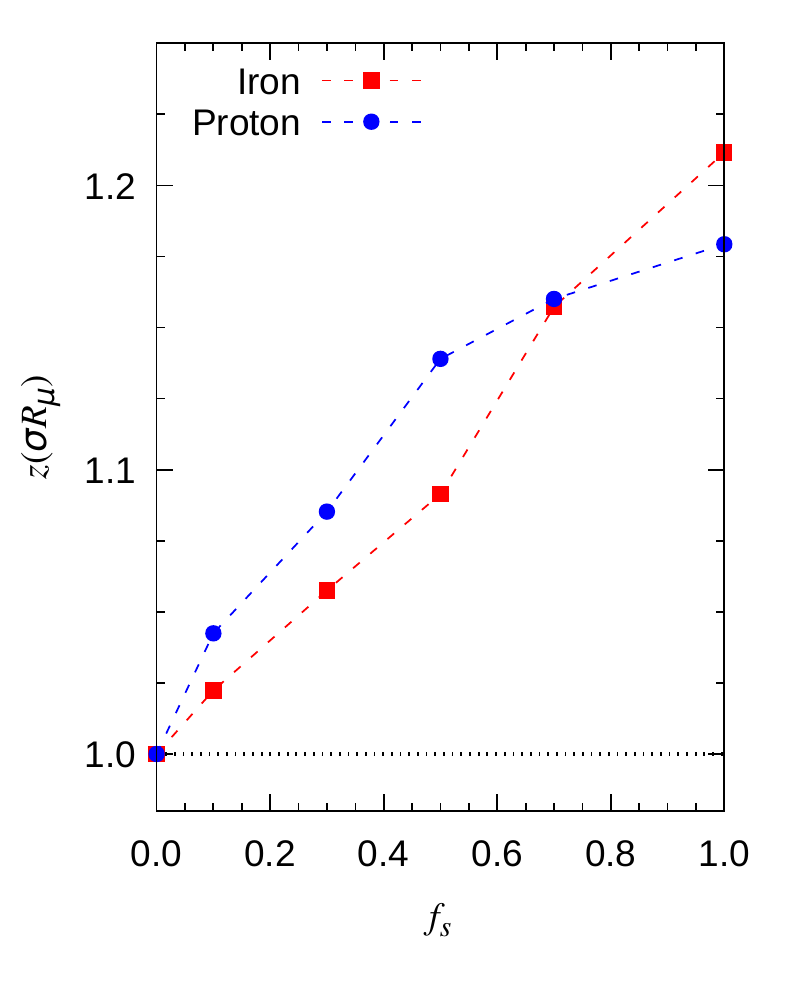}
  \\
  \includegraphics[width=0.4\textwidth]{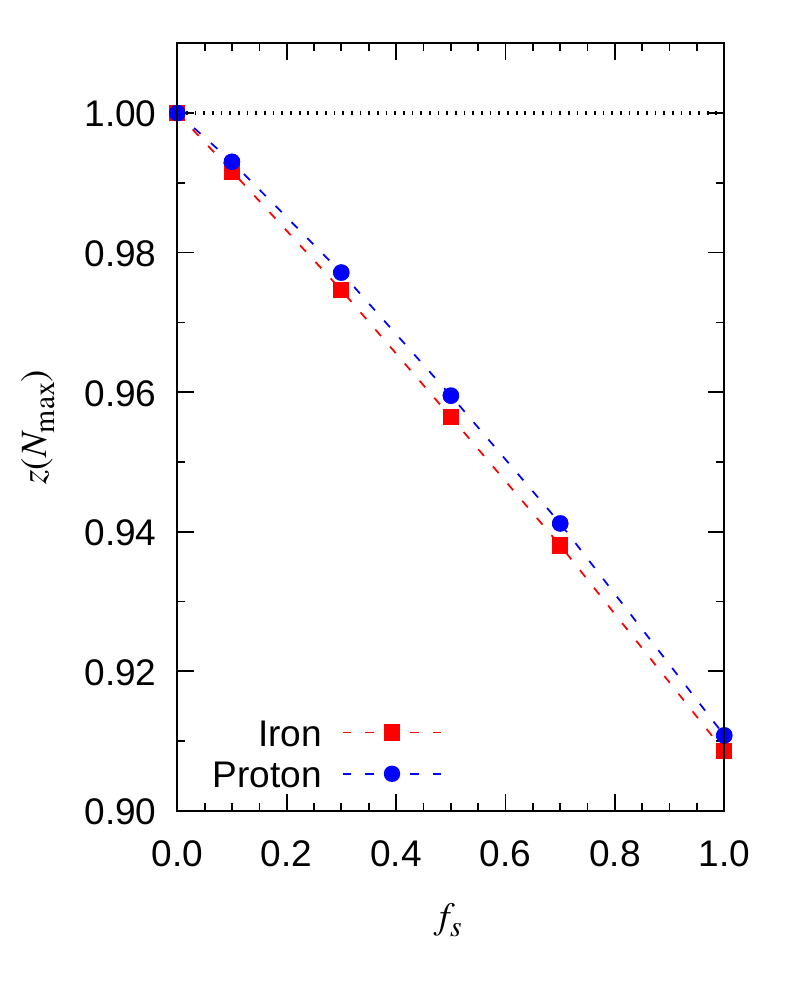}&
  \kern2em &
  \includegraphics[width=0.4\textwidth]{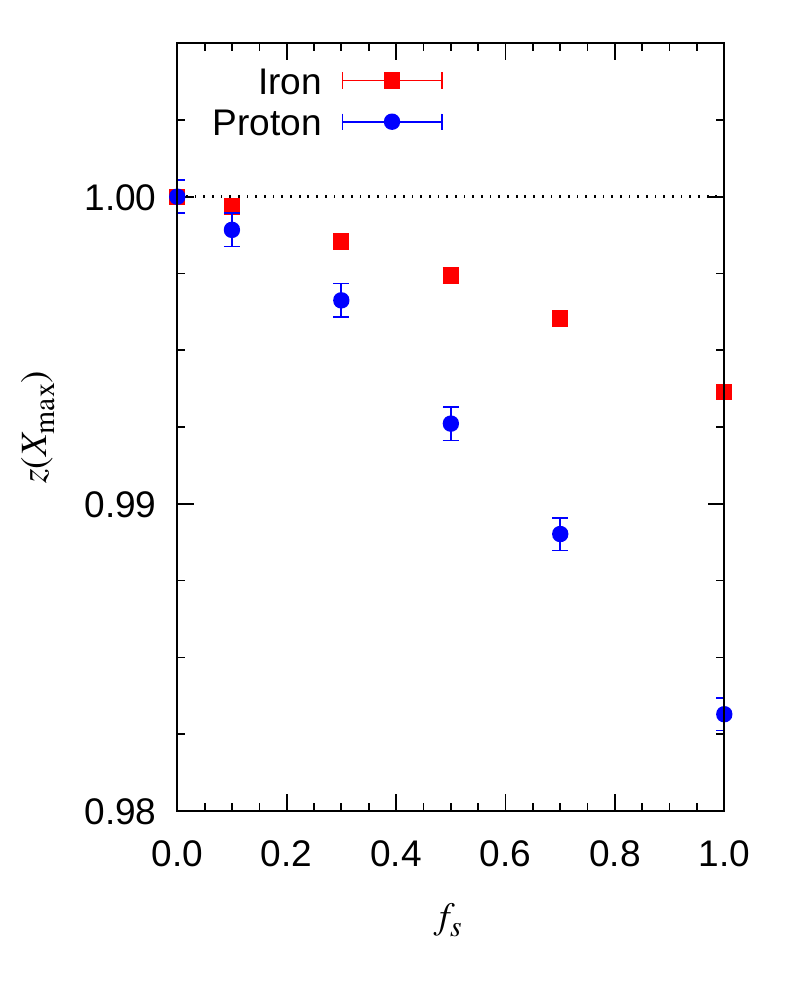}
\\*[-15pt]
\end{tabular}
\caption{$z(R_\mu)$, $z(\sigma R_\mu)$, $z(N_{\rm max})$ and $z(X_{\rm max})$
as a function of $f_s$, for $E_{\rm prim} = 10~{\rm EeV}$, $E_{\rm
smin}=1~{\rm GeV}$, and $E_{\rm pmin} = 1~{\rm PeV}$.  We have run
8000 (20000) showers per point for ground muons (longitudinal
development), setting at each case the thinning algorithm parameters
to get a more detailed simulation of the hadronic or the
electromagnetic cascade, respectively.  \label{fig:3B1}
}
\end{figure*}

\begin{figure*}
\centering
\begin{tabular}{ccc}
  \includegraphics[width=0.4\textwidth]{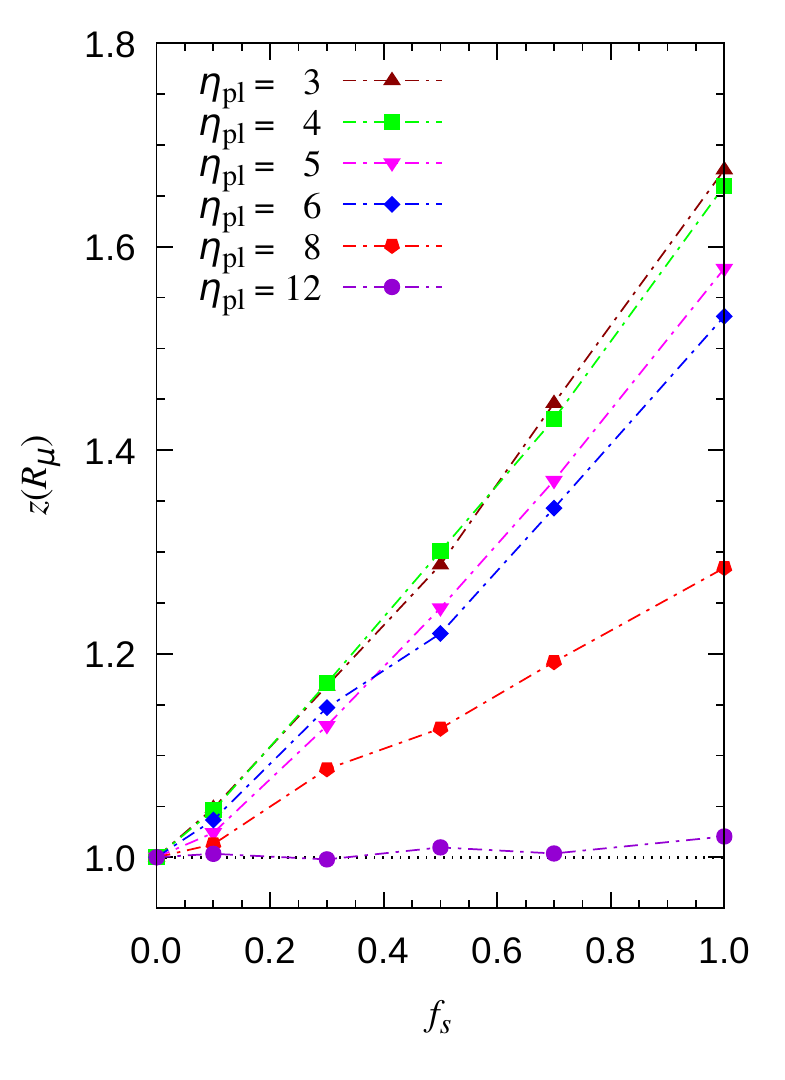}&
  \kern2em &
  \includegraphics[width=0.4\textwidth]{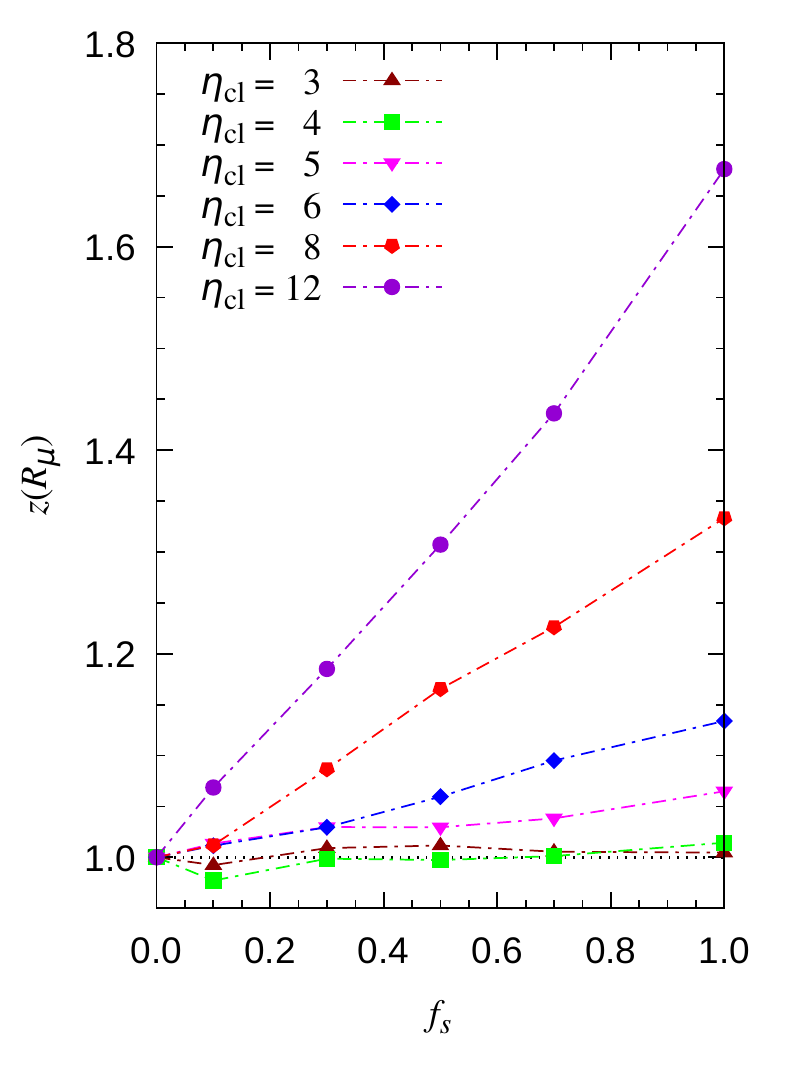}
\\*[-15pt]
\end{tabular}
\caption{$z(R_\mu)$ as a function of $f_s$, with varying 
 limits of the central (cl)  and periferic (pl) regions. The
 figure in the left (right) panel compares the results coming from simulations
 where the swapping algorithm applies to peripheral (central) secondary pions,
 varying the limits of the peripheral (central) region according to the
 functions $F_s^{\rm pl}$ and $F_s^{\rm cl}$,  
 defined in Eqs.~(\ref{Fspl}) and (\ref{Fscl}). \label{fig:3B2}}
\end{figure*}

In Fig.~\ref{fig:3B1} we show $z(R_\mu)$, $z(\sigma
R_\mu)$, $z(N_{\rm max})$, and $z(X_{\rm max})$ as a function of
$f_s$, for $E = 10~{\rm EeV}$, $E_{\rm smin}=1~{\rm GeV}$, and $E_{\rm
pmin} = 1~{\rm PeV}$. We can see that there are no significant changes
with respect to the results shown in Fig.~\ref{fig:dos} for the toy
model. It is remarkable that $\forall f_s$ we have $\sigma R_\mu < R_\mu$, in agreement with Auger observations~\cite{PierreAuger:2021qsd}. In addition, for the fluctuations of $X_{\rm max}$ (not shown in the figure) we reobtain  that $|z(\sigma X_{\rm max})|< 0.03$ for all $f_s\in
[0,1]$. This is because the secondaries emitted in the central
pseudorapidity region have minimal impact on the evolution of the
shower. This is visible in Fig.~\ref{fig:3B2} where we show $z(R_\mu)$
as a function of $f_s$, but with varying limits of the periferic (pl)
and central (cl) regions; namely,
\begin{equation}
F_s^{\rm pl}(\eta_{\rm CM}) =\left\{
\begin{array}{llc}
f_s & \hbox{if} & -\infty < \eta_{\rm CM} < -\eta_{\rm pl} \\*[3mm]
0 & \hbox{if} & -\eta_{\rm pl} \le \eta_{\rm CM} \le \eta_{\rm pl} \\*[3mm]
f_s & \hbox{if} & \eta_{\rm pl} < \eta_{\rm CM} < \infty
\end{array}
\right.
\label{Fspl}
\end{equation}
and
\begin{equation}
F_s^{\rm cl}(\eta_{\rm CM}) =\left\{
\begin{array}{llc}
0 & \hbox{if} & -\infty < \eta_{\rm CM} < -\eta_{\rm cl} \\*[3mm]
f_s & \hbox{if} & -\eta_{\rm cl} \le \eta_{\rm CM} \le \eta_{\rm cl} \\*[3mm]
0 & \hbox{if} & \eta_{\rm cl} < \eta_{\rm CM} < \infty
\end{array}
\right.  \,,
\label{Fscl}
\end{equation}
respectively. Moreover, the plots in Fig.~\ref{fig:3B2}
clearly show that setting $\eta_{\rm pl}=3$ or 4 return virtually the same
results. For $\eta_{\rm pl}>4$, the impact of $\pi\to K$ swapping
diminish with increasing $\eta_{\rm pl}$, as expected, until
presenting a virtually zero impact for $\eta_{\rm
  pl}=12$. Complementary, the curves displayed in the right panel
show that the impact of $\pi\to K$ swapping increases monotonically as
long as the ``central'' region considered gets progressively
wider. For $\eta_{\rm cl} <4$, the central region provides a
negligible contribution to $z(R_\mu)$.

\begin{figure}
\centering
\includegraphics[width=0.45\textwidth]{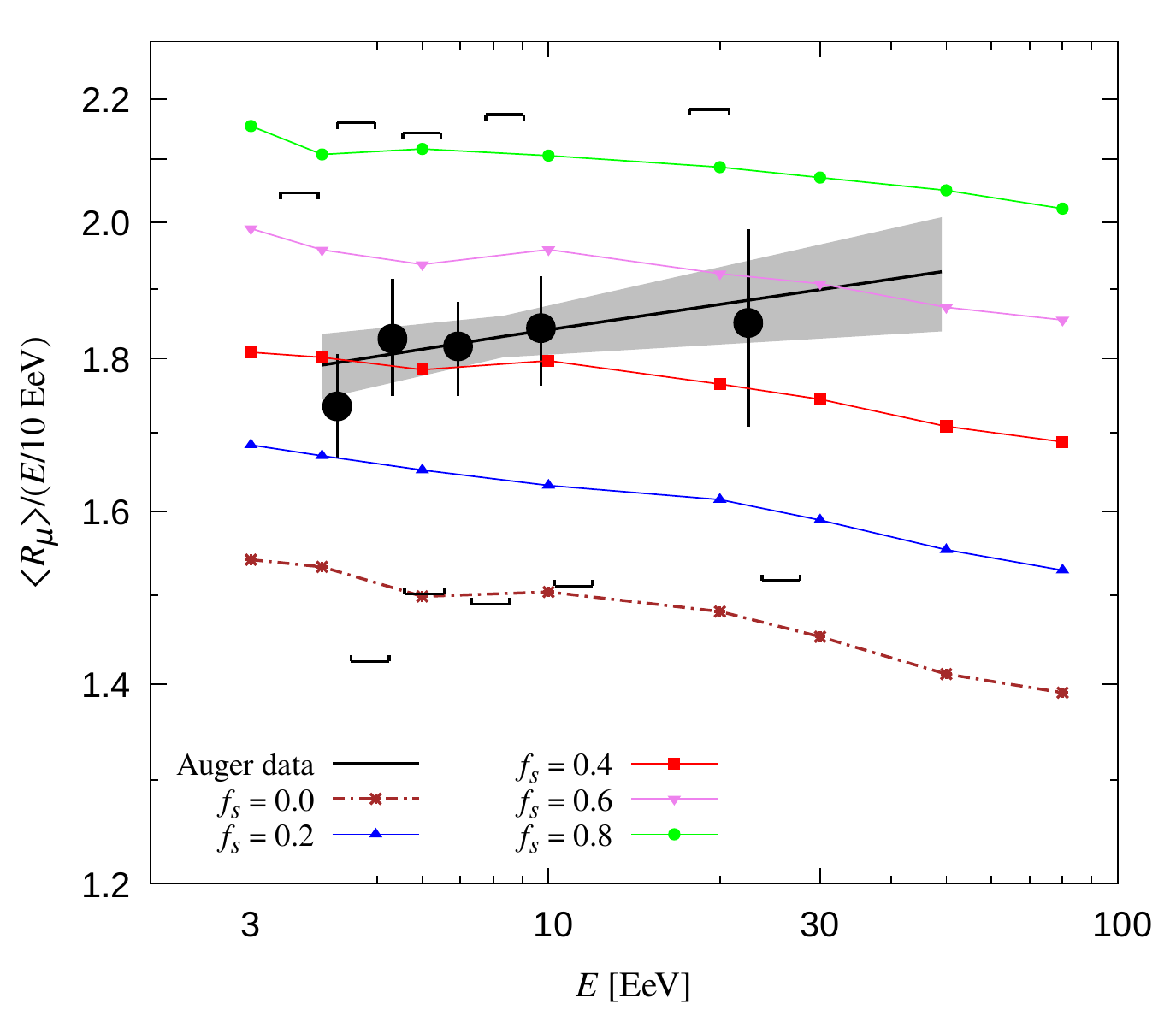}
\caption{Estimations of $R_\mu$ from AIRES simulations for different 
 values of $f_s$ superimposed over Auger data with statistical
 \mbox{($\hspace{0.1em}\bullet\hspace*{-.66em}\mid\hspace{0.16em}$)}
 and systematic (\protect\rotatebox{90}{\hspace{-.075cm}[ ]})
 uncertainties~\cite{Aab:2014pza}. We have adopted the mixed
 baryonic composition shown in the left panel of
 Fig.~\ref{fig:cinco}. \label{fig:3C1}}
\end{figure}

In Fig.~\ref{fig:3C1} we show $\langle R_\mu\rangle/(E/10~{\rm EeV})$
considering the variation of UHECR composition shown in
Fig.~\ref{fig:cinco} and $F_s(\eta_{\rm CM})$ as defined in Eq.~(\ref{Fs}). As expected from the discussion
above, there is no significant differences with the results displayed
in Fig.~\ref{fig:cinco} for the toy
model of Sec.~\ref{sec:2}.  

\def\zs{$\phantom{0}$}\def\spc{$\phantom{,}$\%}\def\lsp{\kern1em\relax}%
\begin{table}
\caption{Global counters for the toy model with $f_s =0.7$, in the case of
$10^{19}\;{\rm eV}$ proton showers inclined $67^\circ$. \label{tabla1}}
\begin{tabular}{lrr}
\hline
\hline
Total hadronic collisions per shower                  &  287,036  &  100.00\spc \\
\lsp Collisions with $E_{\rm proj} < E_{\rm pmin}$          &   284,374  &   99.06\spc \\
\lsp Collisions with $E_{\rm proj} > E_{\rm pmin}$          &     2,662  &     0.94\spc
\\*[5pt]
Total number of secs. produced                         & \zs 7,315,106  & \zs 100.00\spc \\
\lsp Secs. from colls. with $E_{\rm proj} < E_{\rm pmin}$    & 7,036,530  &    96.19\spc \\
\lsp Secs. from colls. with $E_{\rm proj} > E_{\rm pmin}$    &   278,576  &     3.81\spc
\\*[5pt]
Total number of pions scanned                          &   142,550  &     1.95\spc \\
\lsp Pions considered for swapping                     &    56,610  &     0.77\spc \\
\lsp Pions actually swapped                            &    39,609  &     0.54\spc
\\*[2pt]
  \hline
  \hline
\end{tabular}
\end{table}
\begin{table}
  \caption{Global counters for the refined model with $f_s =0.7$, in the case of
$10^{19}\;{\rm eV}$ proton showers inclined $67^\circ$. \label{tabla2}}
\begin{tabular}{lrr}
\hline
\hline
Total hadronic collisions per shower                 &  264,600 & 100.00\spc \\
\lsp Collisions with $E_{\rm proj} < E_{\rm pmin}$         & 262,070 &   99.04\spc \\
\lsp Collisions with $E_{\rm proj} > E_{\rm pmin}$         &   2,530 &    0.96\spc
\\*[5pt]
Total number of secs. produced                       & \zs 6,806,244 & \zs 100.00\spc \\
\lsp Secs. from colls. with $E_{\rm proj} < E_{\rm pmin}$ &     6,544,194 &       96.15\spc \\
\lsp Secs. from colls. with $E_{\rm proj} > E_{\rm pmin}$ &       262,050 &        3.85\spc
\\*[5pt]
Total number of pions scanned                        &       134,060 &        1.97\spc \\
Pions considered for swapping: \\
\lsp Central $(|\eta_{\rm CM}| <4)$                    &        99,790 &        1.47\spc \\
\lsp Peripheral $(|\eta_{\rm CM}| > 4)$                &        34,270 &        0.50\spc \\
\lsp Total (central + peripheral)                    &       134,060 &        1.97\spc \\
\lsp Pions actually swapped                          &        23,988 &        0.35\spc
\\*[2pt]
  \hline
  \hline
\end{tabular}
\end{table}
A few crosschecks on these considerations are in order. In
Tables~\ref{tabla1} and \ref{tabla2} we provide a summary of the
global counters of shower simulations using the toy model and the
refined model, respectively, with $f_s=0.7$. It is interesting to note
that the percentage the pions produced above $E_{\rm pmin}$ remains
the same and is slightly smaller than 2\%. In addition, the number of
collisions and consequently the number of secondaries being produced,
decreases when considering the refined model. This is because in the
toy model we consider secondary neutral pions from the central region
with LAB energy above 1~TeV, and if these pions mutate into kaons they
would most likely interact before decaying, yielding more collisions
in the overal shower and more secondaries. However, the percentage of
the number of pions considered for swapping increases in the
refined model with a ratio of $40\% \div 96\%$. This is because by
lowering the $E_{\rm smin}$ there are many more pions that can be
swapped (some of them with $\eta_{\rm CM} <0$). Looking at the final
figures of pions actually swapped, it shows up that the number of
swapped pions with respect to the number of scanned pions is more or
less the same, and it is actually lower in the refined model; the
ratio is $28\% \div 25\%$. The number of swapped pions when compared
with the number considered for swapping is roughly 70\% in the toy
model and reduces to 27\% in the refined model. Obviously, the ratio
of swapped pions to the {\it effective} number of pions considered for
swapping (i.e., those with $|\eta_{\rm CM} | > 4$) is $f_s =
0.7$. Finally, the number of scanned pions with respect to the total
number of secondaries produced with $E_{\rm proj}> E_{\rm pmin}$ is
roughly 51\%. Note that the fraction of pions produced is larger than
51\% , because in the collisions with $E_{\rm proj} \agt E_{\rm pmin}$
there are several pions that have energy below the threshold.

\section{Sensitivity to $\bm{F_s}$ with LHC Neutrino Experiments}
\label{sec4}

\begin{figure*}[tpb]
\centering
\includegraphics[width=0.90\textwidth]{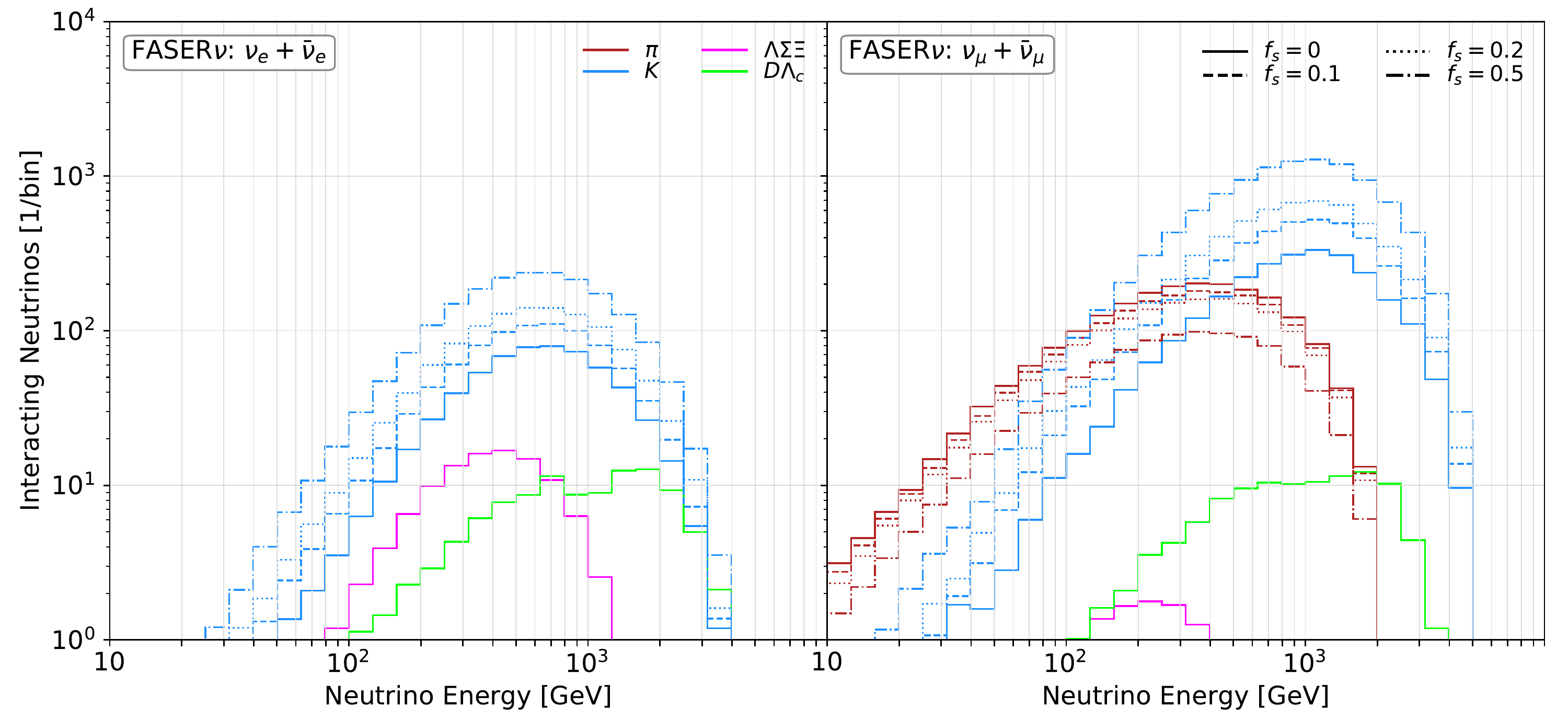}
\caption{Energy spectrum of neutrinos (left) and muon neutrinos 
(right) interacting with FASER$\nu$. The vertical axis shows
the number of charged current neutrino interactions per energy 
bin for an integrated luminosity of $150~{\rm fb}^{-1}$
by different colors: pion decays (red), kaon decays (blue), hyperon
decays (magenta), and charm decays (green). The different line styles
correspond to predictions obtained from SIBYLL-2.3d with secondary pions
processed using the refined model with $F_s(\eta_{\rm CM})$ as in
Eq.~(\ref{Fs}), for different values of $f_s$. 
 \label{fig:FPF1}}
\end{figure*}

\begin{figure*}[tpb]
\includegraphics[width=0.90\textwidth]{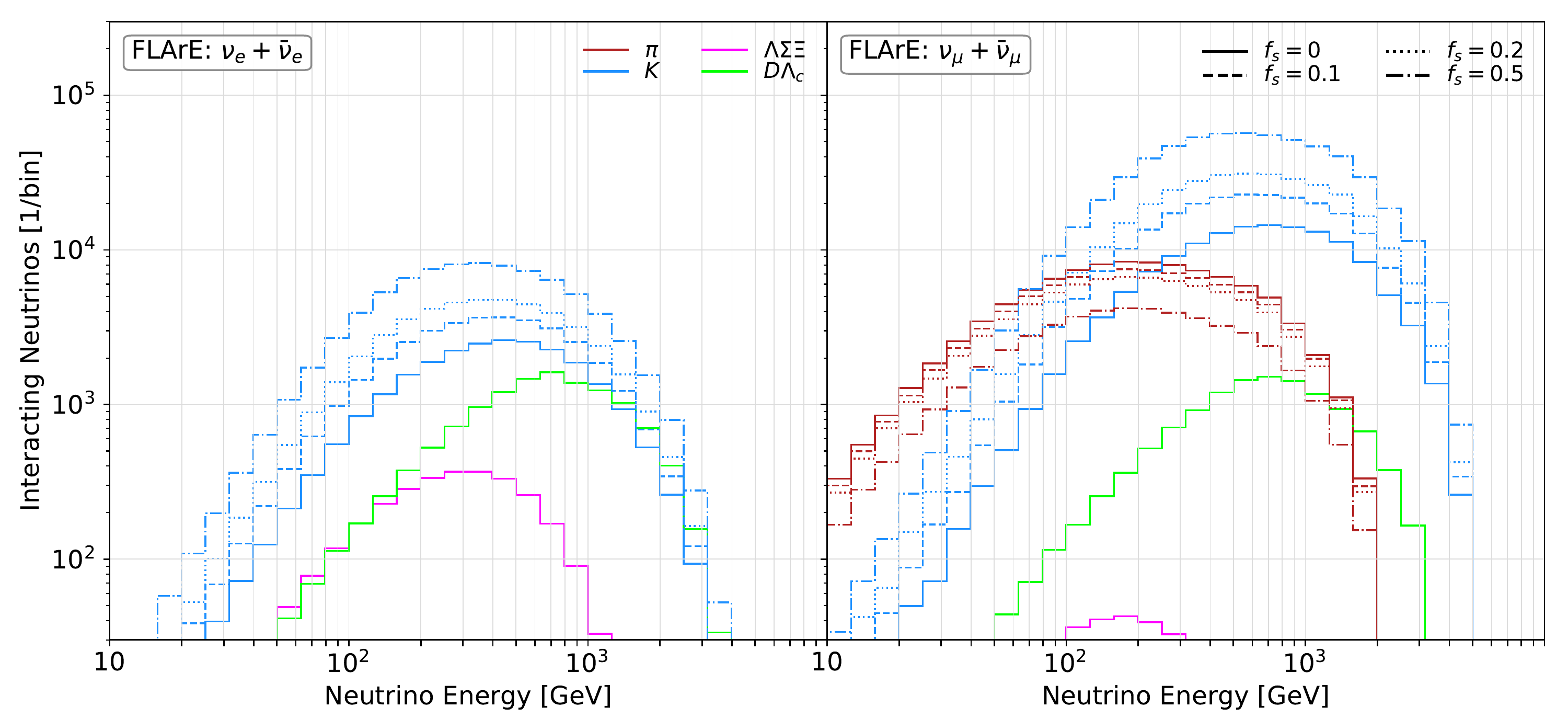}
\caption{Expected number of charged current neutrino interactions
with the FLArE detector at the FPF assuming an integrated luminosity 
of $3~{\rm ab}^{-1}$. See Fig.~\ref{fig:FPF1} for details. 
\label{fig:FPF2}}
\end{figure*}

During the next two decades, the LHC will lengthen the energy 
frontier into both higher energies and much higher luminosities.
Most general-purpose LHC detectors, such as ATLAS, CMS, and 
ALICE are committed to high-$p_T$ physics, featuring events 
with small cross section: ${\cal O}$~(fb, pb, nb). However, 
the total cross section of LHC collisions is ${\cal O}(100~{\rm mb})$. 
Curiously, most of this cross section as well as most of the
highest energy particles produced in these collisions are in the 
far forward region, {\it viz.} at low $p_T$. This implies that 
there is an entire physics program in the far forward region 
which remains to be explored and can indeed be exploited 
during the LHC high luminosity (HL) era. 

One challenge that far-forward detectors in or close to the LHC 
beam pipe have to face are the large particle fluxes and radiation 
levels, essentially restricting their operation to short 
low-luminosity runs. Another possibility is to make use of 
the large flux of LHC neutrinos, which can be probed in 
low-background environments at a safe distance away from 
the interaction point and accelerator infrastructure. Indeed, 
the LHC produces an intense and strongly collimated beam 
of high energy neutrinos in the far-forward direction. These 
neutrinos are mainly produced in the decay of charged pions, 
kaons, hyperons and charmed hadrons, making the measurement 
of the neutrino flux a complimentary probe of forward 
particle production compared to the neutral pion and neutron 
measurements performed at LHCf. 

The feasibility of such LHC neutrino measurements has 
recently been demonstrated by the FASER collaboration,
which reported the observation of the first neutrino 
interaction candidates at the LHC~\cite{FASER:2021mtu}.
Building on this experience, the FASER$\nu$ neutrino 
detector~\cite{FASER:2019dxq, FASER:2020gpr}, which is part 
of the FASER experiment~\cite{Feng:2017uoz, FASER:2018bac}, 
will start its operation already with the LHC Run~3 in 2022. With a target mass of about 1.2~tons 
and an anticipated luminosity of 150~fb$^{-1}$
a total of $\mathcal{O}(10^4)$ muon neutrino and  
$\mathcal{O}(10^3)$ electron neutrino interactions are expected
to be observed. During the HL-LHC, additional far-forward
neutrino experiments have been proposed in the context of the
FPF~\cite{Anchordoqui:2021ghd}. In particular, this includes 
an emulsion based neutrino detector with target mass of about 
20~tons called FASER$\nu$2, a liquid argon based neutrino
detector with target mass of about 10~tons called FLArE and
an electronic neutrino detector called AdvSND. With their 
higher target masses and the HL-LHC luminosity of 
3000~fb$^{-1}$ a large event rate of roughly $10^5$ electron 
neutrino and $10^6$ muon neutrino interactions are expected 
to be observed. 

Both FASER$\nu$ in the near future and the FPF neutrino 
experiments during the HL-LHC would provide a profitable 
arena to measure the pion-to-kaon ratio through the shape of
differential neutrino flux distributions. In particular, 
the pion-to-kaon ratio can be inferred by measuring the 
ratio of electron-to-muon neutrino fluxes. This is because 
pions primarily decay into muon neutrinos, whereas kaon decays
yields a flux of both muon and electron neutrinos. Moreover, 
neutrinos from different parent mesons populate a different 
energy range, and so this can be used to disentangle the fluxes. 
In addition, since $m_\pi < m_K$, neutrinos from pion decay 
are more concentrated around the line-of-sight than those of 
kaon origin, and consequently neutrinos from pions obtain 
less additional transverse momentum than those from kaon decays. 
Hence, the closeness of the neutrinos to the line-of-sight, or
equivalently their rapidity distribution, becomes a compelling 
signal to trace back the neutrino origin to measure the 
pion-to-kaon ratio.

In Fig.~\ref{fig:FPF1}, we show the expected number of neutrino 
interactions with the FASER$\nu$ detector, assuming a
25~cm~$\times$~25~cm cross sectional area and a 1.2~ton target 
mass, as a function of the neutrino energy. Here, we have 
used SIBYLL~2.3d~\cite{Riehn:2019jet} as primary generator and 
use the fast LHC neutrino flux simulation 
introduced Ref~\cite{Kling:2021gos} to describe the propagation 
and decay the long-lived hadrons in the LHC beam pipe. The origin of
the neutrinos is indicated by the different line colors: red for 
pion decay, blue for kaon decay, magenta for hyperon decay, 
and green for charm decay. As explained above, the neutrinos from pions
and kaons populate different regions of phase space, which can be
used to disentangle pion and kaon production. In Fig.~\ref{fig:FPF2}, 
we also show the results for the FLArE detector at the FPF, which 
is assumed to have a 1~m$~\times$~1~m cross sectional area and a 10~ton 
target mass. 

In Fig.~\ref{fig:FPF1} and Fig.~\ref{fig:FPF2}, we also show how a 
$\pi\leftrightarrow K$ swapping as defined in Eq.~(\ref{Fs}) changes
the expected neutrino fluxes and event rates for the considered
experiments. As expected,  positive values of $f_s$ lead to a 
suppression of the neutrino flux from pions as well as a larger 
relative enhancement of the neutrino flux from kaons. This is due 
to the  initially roughly 10 times larger flux of pions, 
such that even a small rate of $\pi \leftrightarrow K$ swapping 
can substantially increase the neutrino flux from the kaon decays. 
This leads to the remarkable result that already for $f_s=0.1$ 
($f_s=0.2$) the predicted electron neutrino flux at the peak of the 
spectrum is a factor of 1.6 (2.2) larger. These differences are 
significantly larger than the anticipated statistical uncertainties 
at the FPF~\cite{Kling:2021gos, Anchordoqui:2021ghd}. This let's 
us conclude that LHC neutrino flux measurments with new forward 
detectors at the LHC will provide invaluable complementary information 
to test our model and its improvements, together with eventual 
alternative ones, addressing the muon puzzle via strangeness 
enhancement.
\begin{figure*}
\begin{center}
\begin{tabular}{ccc}
\kern7mm{\Large\bf (a)} && \kern7mm{\Large\bf (b)}\\*[-1mm]
\includegraphics[width=0.4\textwidth]{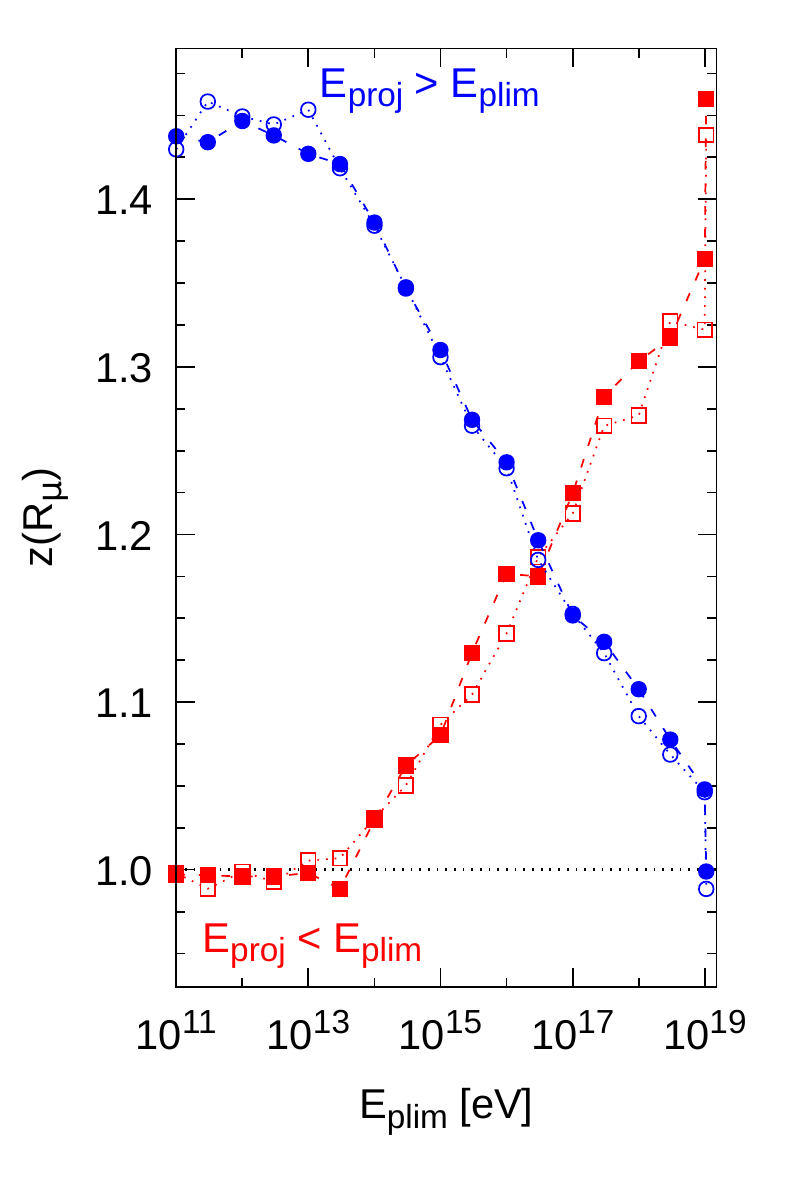}&
\kern3em &
\includegraphics[width=0.4\textwidth]{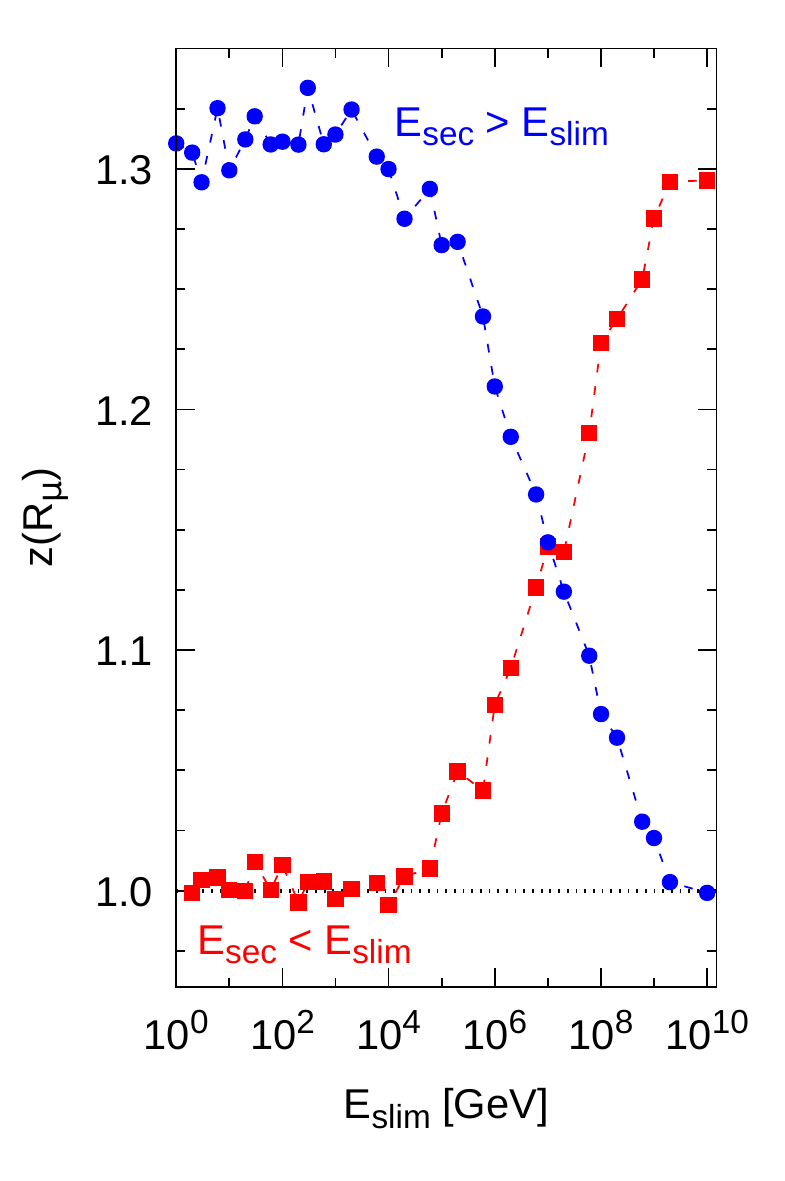}
\end{tabular}
\end{center}
\caption{$z(R_\mu)$ as a function of $E_{\rm plim}$ (a) and $E_{\rm
    slim}$
  (b) for fixed $f_s =0.5$. {\bf (a) Varying projectile energy range, with fixed secondary energy range.}  $[E_{\rm smin},E_{\rm smax}]$ is always kept fixed
  and equal to $[1\;{\rm TeV},\infty]$ (solid symbols) or $[1\;{\rm
      GeV},\infty]$ (open symbols). Each blue
    circle (red square) in the figure 
  corresponds to simulations run with $[E_{\rm pmin},E_{\rm
      pmax}]=[E_{\rm plim},\infty]$ ($[E_{\rm pmin},E_{\rm
      pmax}]=[90\;{\rm GeV},E_{\rm plim}]$), $100\;{\rm GeV}\le E_{\rm plim}
   \le 10.05\;{\rm EeV}$. {\bf (b) Fixed projectile energy range, with
     varying secondary energy range.}  $[E_{\rm pmin},E_{\rm pmax}]$ is always kept fixed
  and equal to $[1\;{\rm PeV},\infty]$.
  Each blue circle (red square)
  corresponds to simulations run with $[E_{\rm smin},E_{\rm
      smax}]=[E_{\rm slim},\infty]$ ($[E_{\rm smin},E_{\rm
      smax}]=[1\;{\rm GeV},E_{\rm slim}]$), $1\;{\rm GeV}\le E_{\rm slim}
   \le 10\;{\rm EeV}$. \label{fig:app1}}
\end{figure*}
\begin{figure*}
\centering
\begin{tabular}{ccc}
\includegraphics[width=0.4\textwidth]{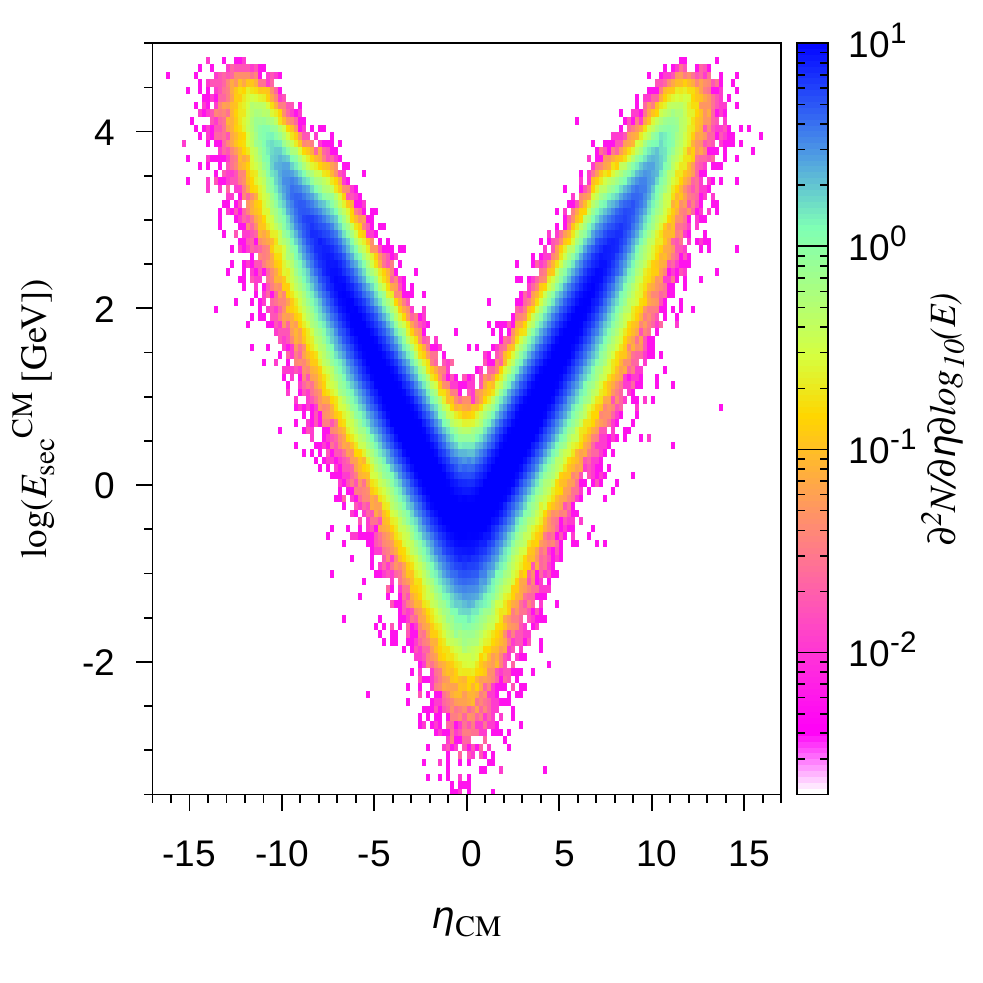}&
\kern3em &
\includegraphics[width=0.4\textwidth]{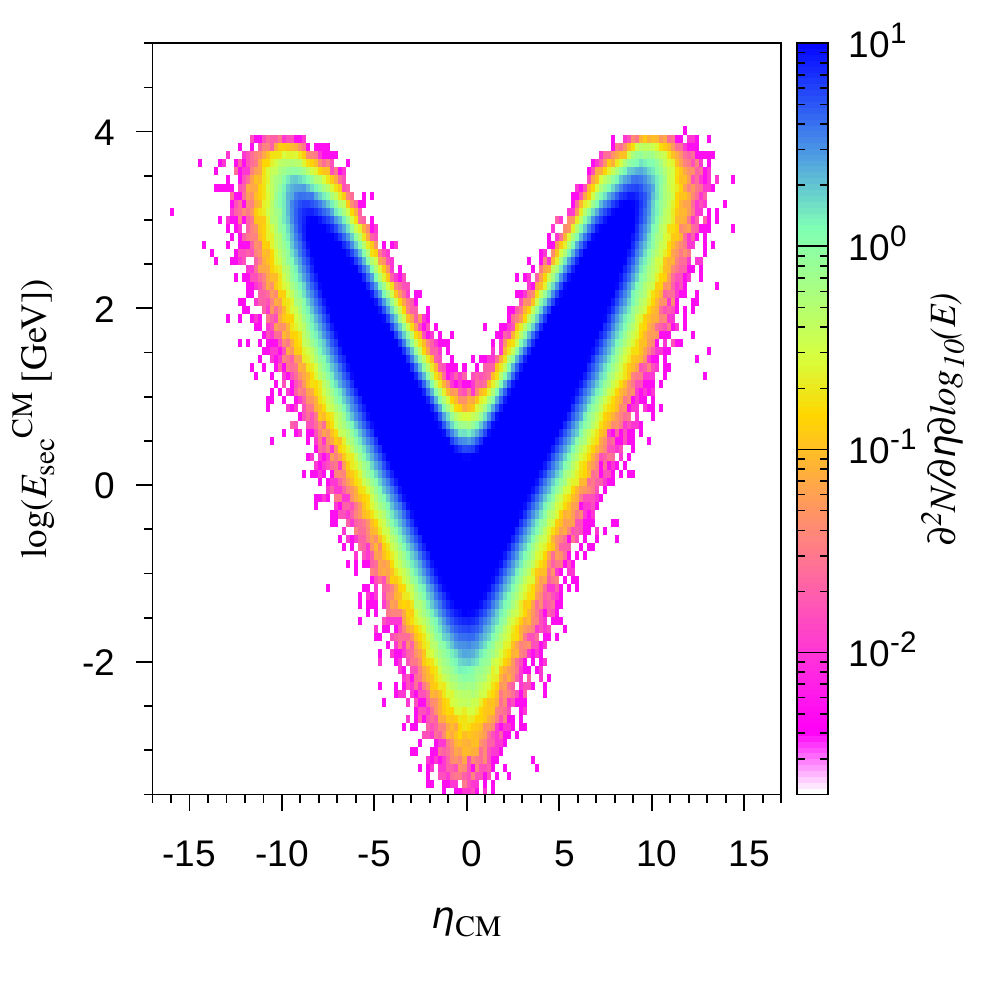}
\\
\includegraphics[width=0.4\textwidth]{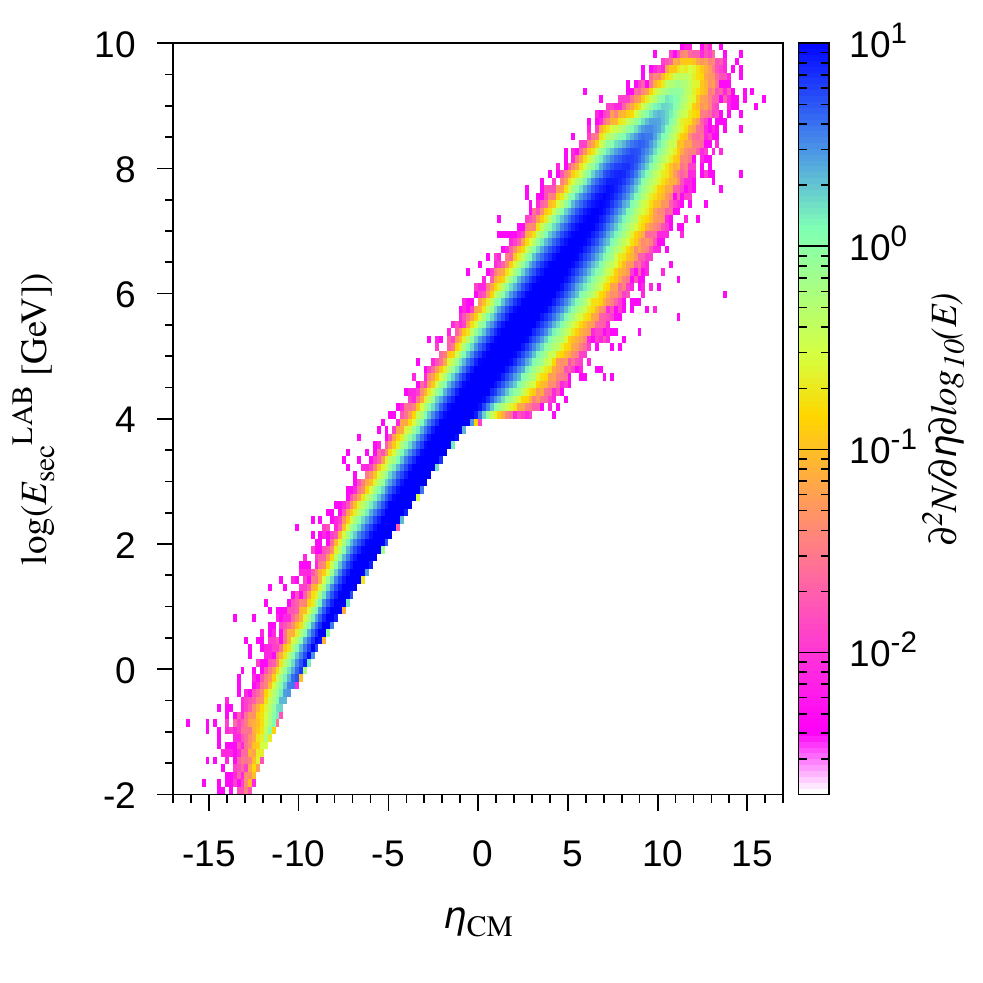}&
\kern3em &
\includegraphics[width=0.4\textwidth]{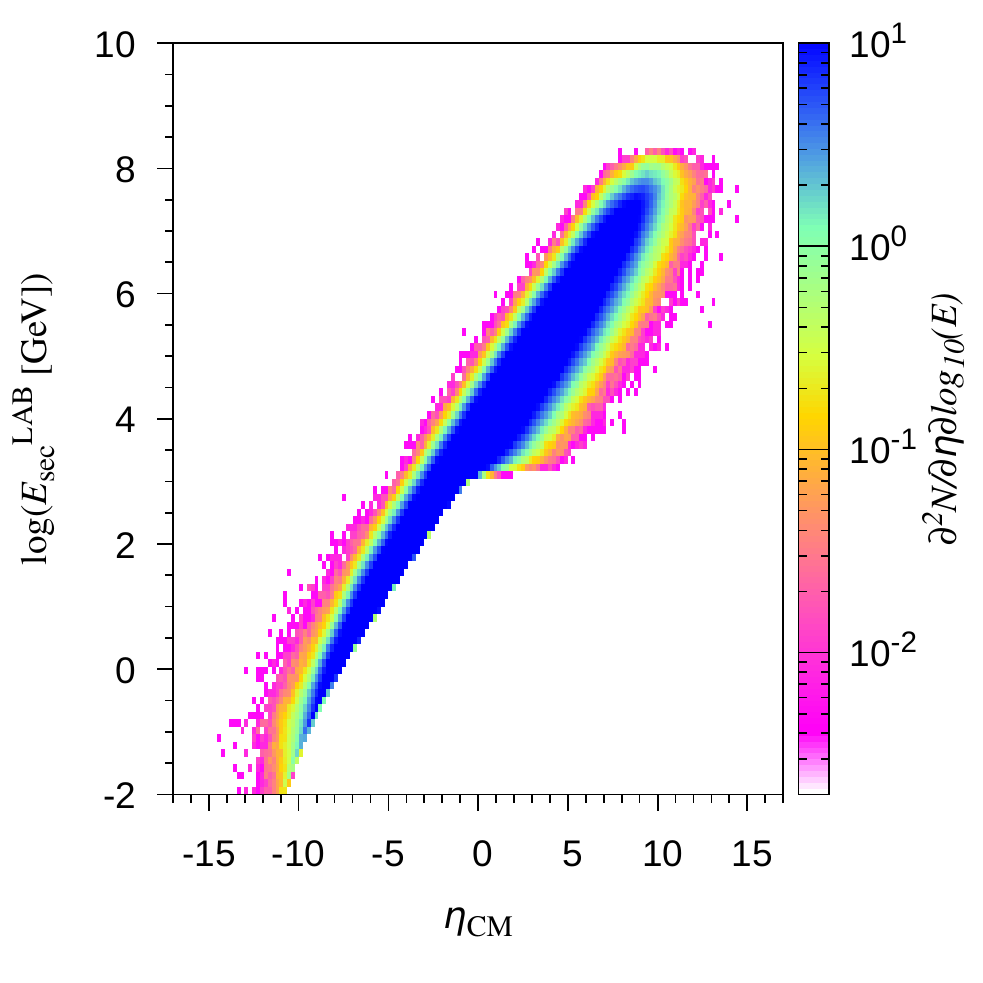}
\end{tabular}
\caption{Pion $E_{\rm sec}^{\rm CM}$ vs $\eta_{\rm CM}$ (upper row)
and $E_{\rm sec}^{\rm LAB}$ vs $\eta_{\rm CM}$ (lower row) bivariate
  distributions. The left (right) column correspond to the results from $10^4$
  collisions of a 10 EeV proton (iron nucleus) scattering
  off a proton (nitrogen nucleus) at
  rest, simulated with EPOS-LCH 1909. \label{fig:appB12}}
\end{figure*}

\section{Conclusions}
\label{sec5}

We have examined the influence of $\pi \leftrightarrow K$ swapping on
the development of extensive air showers. We constructed an empirical testable 
model, based on ALICE observations of the enhancement of strangeness
production in high-energy hadronic collisions, which can accommodate
the muon deficit between simulations and Auger data.\footnote{One possible realization of our phenomenological model may be obtained by considering collective statistical hadronization effects into the standard string fragmentation process~\cite{Baur:2019cpv}.} We derived a
parametrization of the $\pi \leftrightarrow K$ swapping probability in
terms of the pseudorapidity and the nucleus baryon number.

We have also explored potential strategies for model improvement using
the massive amounts of data to be collected at the FASER$\nu$ and future LHC neutrino experiments at the  FPF. We have shown
that these experiments will attain sensitivity to probe the model phase space.

Within this decade, ongoing detector upgrades of existing facilities,
such as AugerPrime~\cite{PierreAuger:2016qzd} and
IceCube-Gen2~\cite{IceCube-Gen2:2020qha}, will enhance the precision
of air shower measurements and reduce uncertainties in the
interpretation of muon data. In particular, as a part of the upcoming
AugerPrime upgrade each surface station will have additional detectors
that will provide complementary measurements of the incoming shower
particles, consequently leading to improved reconstruction of muons
and electromagnetic particles~\cite{PierreAuger:2016qzd}. This will
allow for the measurement of the properties of extensive air showers
initiated by the highest energy cosmic rays with unprecedented
precision. As we have shown in this paper, future Auger measurements
will be highly complemented by observations at the LHC neutrino experiments which will
provide a unique determination of the pion-to-kaon ratio at LHC
energies.  Altogether this will provide a powerful test of models
addressing the muon puzzle via strangeness enhancement.

\section*{Acknowledgements}
We thank our colleagues from the Pierre Auger Collaboration for 
valuable discussion. L.A.A. and J.F.S. are supported by U.S. 
National Science Foundation (NSF Grant PHY-2112527). C.G.C. and 
S.J.S. are partially supported by ANPCyT. The work of F.K. is 
supported by the Deutsche Forschungsgemeinschaft under Germany’s
Excellence Strategy - EXC 2121 Quantum Universe - 390833306.

\appendix

\section{Limitting Projectile and Secondary Energies}
\label{appA}

In this Appendix we analyze the variation of
$z(R_\mu)$ with both projectile and secondary energies for fixed $f_s$. To this end we introduce the new
variables $E_{\rm plim}$ and $E_{\rm slim}$ to limit the maximum and
minimum energies of the projectile $E_{\rm proj}$ and secondary $E_{\rm sec}$,
respectively.  In Fig.~\ref{fig:app1} we show  $z(R_\mu)$ as a function of $E_{\rm plim}$ and $E_{\rm
    slim}$, for fixed $f_s =0.5$. By analyzing the variation of $z(R_\mu)$ with
  $E_{\rm plim}$ and $E_{\rm slim}$ we conclude that:
\begin{itemize}[noitemsep,topsep=0pt]
\item The impact of the substitution of $\pi$'s by $K$'s reaches a
  maximum when $0<E_{\rm pmin} \alt 10~{\rm TeV}$.
\item In (a), at $E_{\rm plim}\simeq 10^{19}\;{\rm eV}$, both the blue
  and red sets show pairs of points significantly apart: they
  correspond to values of $E_{\rm plim}$ slightly smaller or larger
  than the primary energy ($10^{19}\;{\rm eV}$), that respectively
  prevents or not the application of the swapping algorithm to the
  first hadronic interaction at the beginning of the shower
  development. This reveals that the first interaction has, by itself,
  a finite impact of the final number of muons at ground.
\item There are no significant differences between the open and solid
  symbols plots included in (a). This means that swapping of low
  energy pions ($E_{\rm sec}$ lower than 1 TeV) has no visible impact
  on $z(R_\mu)$. This also shows up clearly in (b) where the blue
  points remain around the maximum value for $E_{\rm slim}\alt
  1\;{\rm TeV}$.
\end{itemize}

\section{EPOS-LHC}
\label{appB}

In this Appendix we report on the results of simulated particle
collisions with EPOS-LHC~\cite{Pierog:2013ria}.  In
Fig.~\ref{fig:appB12} we show bivariate distributions of secondary
pions. From a comparison with
Fig.~\ref{fig:3A12} we see that there are no major
differences in the distributions, but just a small deviation of the
predicted multiplicity in the central region.

We have shown elsewhere~\cite{Sciutto:2019pqs} that the discrepancy between Auger data
and air shower simulations with SIBYLL 2.3d is slightly smaller than
the discrepancy obtained from simulations with EPOS-LHC 1909. For
showers process with QGSJetII-04 hadronic event generator~\cite{Ostapchenko:2010vb}, the
discrepancy between data and simulations is even larger~\cite{Aab:2016hkv}.  This
justifies the choice of SIBYLL 2.3d in our study.

\end{document}